\pdfoutput=1
\documentclass[useAMS,usenatbib]{mn2e}
\usepackage{graphicx}
\usepackage{amsmath}
\usepackage{txfonts}
\usepackage{amssymb}
\usepackage{booktabs}

\title[CANDELS: host galaxies of z $\sim$ 0.7 AGN]{Morphologies of z$\sim$0.7 AGN Host Galaxies in CANDELS: No trend of merger incidence with AGN luminosity}
\author[Villforth et
al.]{C. Villforth$^{1,2}$, F. Hamann$^{1}$, D. J. Rosario$^{3}$, P. Santini$^{4}$, E. J. McGrath$^{5}$, A. van der Wel $^{6}$, \newauthor Y. Chang $^{6}$, Y. Guo$^{7}$, T. Dahlen$^{8}$, E. F. Bell$^{9}$, C. J. Conselice$^{10}$, D. Croton$^{11}$, A. Dekel$^{12}$,  \newauthor S. M. Faber$^{7}$, N. Grogin$^{8}$, T. Hamilton$^{13}$, P. F. Hopkins$^{14,15}$, S. Juneau$^{16}$, J. Kartaltepe$^{17}$, \newauthor  D. Kocevski$^{18}$, A. Koekemoer$^{8}$, D. Koo$^{7}$, J. Lotz$^{8}$, D. McIntosh$^{19}$, M. Mozena$^{7}$, R. Somerville$^{20}$, \newauthor  V. Wild$^{2}$\\
$^{1}$Department of Astronomy, University of Florida, 32611 Gainesville, FL,
USA\\
$^{2}$ SUPA, University of St. Andrews, School of Physics and Astronomy, North Haugh, KY16 9SS, St. Andrews, Fife, Scotland\\
$^{3}$Max-Planck-Institut f\"{u}r Extraterrestrische Physik (MPE), Postfach 1312, 85741 Garching, Germany\\
$^{4}$INAF – Osservatorio Astronomico di Roma, via di Frascati 33, 00040 Monte Porzio Catone, Italy\\
$^{5}$Department of Physics and Astronomy, Colby College, Waterville, ME 04901, USA\\
$^{6}$Max Planck Institute for Astronomy, K\"{o}nigstuhl 17, 69117 Heidelberg, Germany\\
$^{7}$University of California Observatories/Lick Observatory, University of California, Santa Cruz, CA 95064, USA\\
$^{8}$Space Telescope Science Institute, 3700 San Martin Dr., Baltimore, MD, 21218, USA\\
$^{9}$Department of Astronomy, University of Michigan, Ann Arbor, MI, 48104, USA\\
$^{10}$University of Nottingham, School of Physics and Astronomy, Nottingham NG7 2RD, UK\\
$^{11}$Centre for Astrophysics and Supercomputing, Swinburne University of Technology, PO Box 218, Hawthorn, VIC 3122, Australia\\
$^{12}$Racah Institute of Physics, The Hebrew University, Jerusalem 91904, Israel\\
$^{13}$Department of Natural Sciences, Shawnee State University, Portsmouth, OH 45662, USA\\
$^{14}$TAPIR, Mailcode 350-17, California Institute of Technology, Pasadena, CA 91125, USA\\
$^{15}$Department of Astronomy and Theoretical Astrophysics Center, University of California Berkeley, Berkeley, CA 94720, USA\\
$^{16}$CEA Saclay, IRFU/SAp, 91191 Gif-Sur-Yvette, France\\
$^{17}$National Optical Astronomy Observatory, 950 N. Cherry Ave., Tucson, AZ 85719, USA\\
$^{18}$Department of Physics and Astronomy, University of Kentucky, Lexington, KY 40506, USA\\
$^{19}$Department of Physics \& Astronomy, University of Missouri-Kansas City, 5110 Rockhill Road, Kansas City, MO 64110, USA\\
$^{20}$Department of Physics and Astronomy, Rutgers University, 136 Frelinghuysen Road, Piscataway, NJ 08854, USA
}
\begin{document}

\date{}

\pagerange{\pageref{firstpage}--\pageref{lastpage}} \pubyear{2014}

\maketitle

\label{firstpage}
\clearpage
\begin{abstract}
The processes that trigger Active Galactic Nuclei (AGN) remain poorly understood. While lower luminosity AGN may be triggered by minor disturbances to the host galaxy, stronger disturbances are likely required to trigger luminous AGN. Major wet mergers of galaxies are ideal environments for AGN triggering since they provide large gas supplies and galaxy scale torques. There is however little observational evidence for a strong connection between AGN and major mergers. We analyse the morphological properties of AGN host galaxies as a function of AGN and host galaxy luminosity and compare them to a carefully matched sample of control galaxies. AGN are X-ray selected in the redshift range $0.5 < z < 0.8$ and have luminosities $41 \la \log(L_X \ \textrm{[erg/s]}) \la 44.5$. `Fake AGN' are simulated in the control galaxies by adding point sources with the magnitude of the matched AGN. We find that AGN host and control galaxies have comparable assymetries, Sersic indices and ellipticities at restframe $\sim$950nm. AGN host galaxies show neither higher average asymmetries nor higher fractions of very disturbed objects. There is no increase in the prevalence of merger signatures with AGN luminosity. At 95\% confidence we find that major mergers are responsible for $<$6\% of all AGN in our sample as well as $<$40\% of the highest luminosity AGN ($\log(L_X \ \textrm{[erg/s]}) \sim 43.5$). Major mergers therefore either play only a very minor role in the triggering of AGN in the luminosity range studied or time delays are too long for merger features to remain visible.
\end{abstract}


\section{Introduction}
\label{S:intro}

Supermassive black holes (SMBHs) are now believed to be present in the centres of most if not all massive galaxies \citep[e.g.][and references therein]{kormendy_coevolution_2013}. While most SMBHs do not accrete large amounts of gas, a small fraction of them show strong signs of accretion. These objects are known as Active Galactic Nuclei (AGN). The conditions under which SMBHs become active remain poorly understood. AGN activity requires a) the availability of either gas or stars to feed the black hole and b) a process to strip said material of its angular momentum. Depositing large amounts of gas in the centres of galaxies makes AGN activity probable since it provides material, as well as an ideal environment to transfer angular momentum.

Different processes could provide such a favourable environment for AGN triggering: major and minor mergers of galaxies \citep[e.g.][]{silk_quasars_1998,hopkins_cosmological_2008,di_matteo_direct_2008,springel_black_2005}, bars \citep[e.g.][]{shlosman_bars_1989}, close passages of galaxies disturbing the gravitational potential \citep[e.g.][]{hopkins_cosmological_2008}, cooling of gas from the hot halo \citep[e.g.][]{croton_many_2006,pope_investigating_2012,fabian_observational_2012}, mass loss from stellar winds \citep[e.g.][]{davies_stellar_2012}, cold flows in combination with violent disk instabilities \citep{dekel_cold_2009,bournaud_black_2011}, galaxy scale torques \citep{angles-alcazar_black_2013} as well as accretion of small amounts of gas from the halo \citep{king_fuelling_2007}.

AGN of different luminosities require vastly different amounts of accretion material. Given typical AGN lifetimes of $10^{8}$yr \citep[see e.g.][and references therein, although a wide range of AGN lifetimes is possible]{martini_quasar_2001,yu_observational_2002,martini_qso_2003} a low luminosity AGN of $\log(L_{bol} \ \textrm{[erg/s]})=42$ requires as little as $2\times 10^{4} M_{\odot}$, while a luminous AGN of $\log(L_{bol} \ \textrm{[erg/s]})=46$ requires as much as $2\times 10^{8} M_{\odot}$. This corresponds to about 1\% of the total mass in typical massive galaxies \citep{catinella_galex_2010,saintonge_cold_2011}. Stripping such a substantial fraction of the gas mass in a galaxy of large parts of its angular momentum in the short lifetime of the AGN is challenging. It is hence expected that while a wide range of processes can lead to the triggering of low luminosity AGN, high luminosity AGN require substantial disturbances to their host galaxies. Major mergers of galaxies are therefore thought to dominate triggering at the highest AGN luminosities. While there is no clear predicted break point, a transition is likely around $\log(L_{bol} \ \textrm{[erg/s]})=46$, as argued above \citep[see also][]{hopkins_we_2013}.

The topic of AGN triggering became relevant for galaxy evolution when it was discovered that super-massive black holes are not only common in massive galaxies but their masses also correlate well with the properties of their host galaxies(velocity dispersion \citep[e.g.][]{gebhardt_relationship_2000,ferrarese_fundamental_2000,tremaine_slope_2002,gultekin_m-_2009,gultekin_observational_2011}, stellar mass \citep[e.g.][]{haring_black_2004}, central light concentration \citep{graham_correlation_2001} as well as absolute magnitudes in some bands \citep{mclure_black_2002,marconi_relation_2003}). Theoretical models posit that major mergers of gas-rich galaxies trigger both starbursts and AGN, and the AGN subsequently shuts down the star formation by depositing energy into the ISM (commonly termed AGN feedback) and thereby establishes the $M-\sigma$ relation \citep{hopkins_cosmological_2008,somerville_semi-analytic_2008,king_black_2003,silk_quasars_1998,di_matteo_energy_2005,springel_black_2005}. Although some authors have pointed out that repeated mergers of galaxies containing black hole seeds explain the correlations as well \citep{peng_how_2007,jahnke_non-causal_2010,angles-alcazar_black_2013}.

Despite its high theoretical appeal, observational evidence for a connection between mergers and AGN remains mixed. For certain samples of AGN, rates of recent mergers are extremely high. This is true for local quasars that also show large FIR luminosities - similar to ULIRGs - as well as peculiar AGN samples such as red quasars \citep{canalizo_quasi-stellar_2001,urrutia_evidence_2008}. However, when host galaxies of AGN are compared to galaxies of similar mass they show comparable incidences of disturbances indicative of recent mergers \citep{kocevski_candels:_2012,cisternas_bulk_2011,dunlop_quasars_2003,boehm_agn_2012,grogin_agn_2005,gabor_active_2009,pierce_host_2010,pierce_aegis:_2007}. However, host galaxies of moderately luminous AGN and radio-selected AGN show weak merger features with higher surface brightnesses than found in control galaxies \citep{bennert_evidence_2008,ramos_almeida_are_2011}. \citet{treister_major_2012} studied the incidence of merger features as a function of AGN luminosity and found that the highest luminosity AGN have higher incidences of mergers. However, the sample was not uniformly selected and no control samples were used. It is therefore unclear if these results will hold in well-controlled studies.

In this study, we examine the incidence of disturbances indicative of major mergers in AGN hosts as a function of AGN luminosity compared to a carefully matched control sample using CANDELS \citep{koekemoer_candels:_2011,grogin_candels:_2011} data in GOODS-S. One particular goal of this study is to test the hypothesis that mergers dominate triggering in more luminous AGN. We select a sample of AGN spanning a wide range in luminosities using their X-ray emission only. The most luminous AGN in our sample require accreting material around $10^{8} M_{\odot}$ assuming standard quasar lifetimes. We aim to answer the question of whether merger triggering becomes more important with increasing AGN luminosity and, if so, at what AGN luminosity mergers become the dominant mechanism. The incidence of major mergers will be assessed using quantitative morphological measures that determine the level of disturbance in the host galaxy \citep{conselice_asymmetry_2000}. While many studies rely on human classifiers \citep[e.g.][]{kocevski_candels:_2012,cisternas_bulk_2011,treister_major_2012}, using quantitative measures enables us to detect more subtle levels of disturbance \citep{conselice_asymmetry_2000,lotz_effect_2010,lotz_effect_2010-1} as well as allow more detailed statistical analysis of the results. 

The paper is organized as follows: data reduction and analysis as well as sample selection are presented in Section \ref{S:data}. The results are presented in Section \ref{S:results}, followed by discussion in Section \ref{S:discussion} and summary and conclusion in Section \ref{S:summary}. Supplemental information about morphological measures used and simulations performed are presented in Appendix \ref{S:appendix}. The cosmology used is $H_{0}=70\textrm{km s}^{-1}\textrm{Mpc}^{-1},\Omega_{\Lambda}=0.7, \Omega_{m}=0.3$. Throughout the paper, we use AB magnitudes.

\section{Data \& Analysis}
\label{S:data}

For the study, we use the $F160W/H$ band imaging data in GOODS-South \citep{grogin_candels:_2011}. The data reduction is described in detail in \citet{koekemoer_candels:_2011}. Throughout the paper, when citing magnitudes, we refer to the observed frame F160W band, which corresponds to a rest-frame wavelength of $\sim$950nm in the redshift range studied.

\subsection{AGN Sample}
\label{S:sample}

\begin{figure*}
\begin{center}
\includegraphics[width=8.5cm]{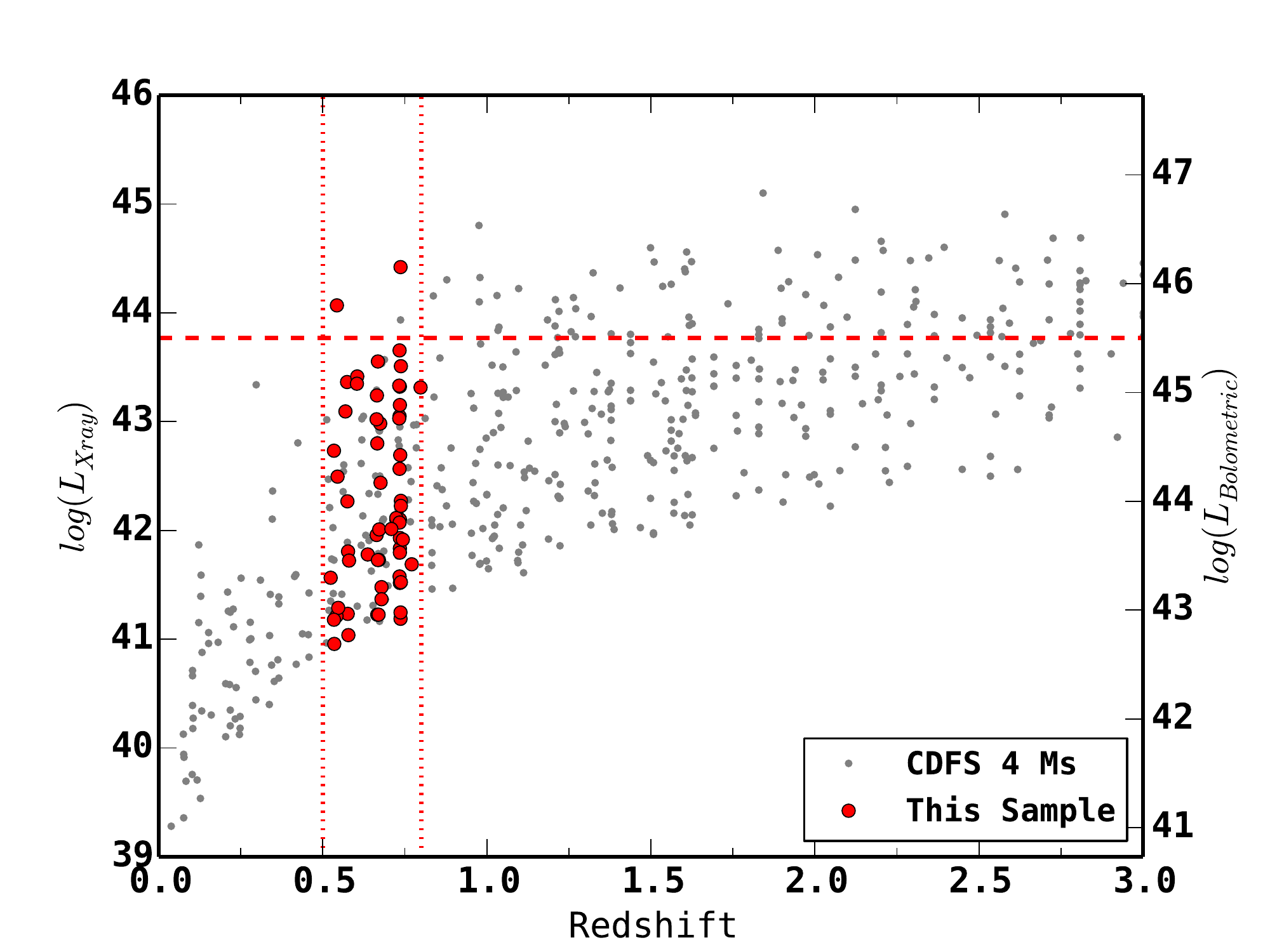}
\includegraphics[width=8.5cm]{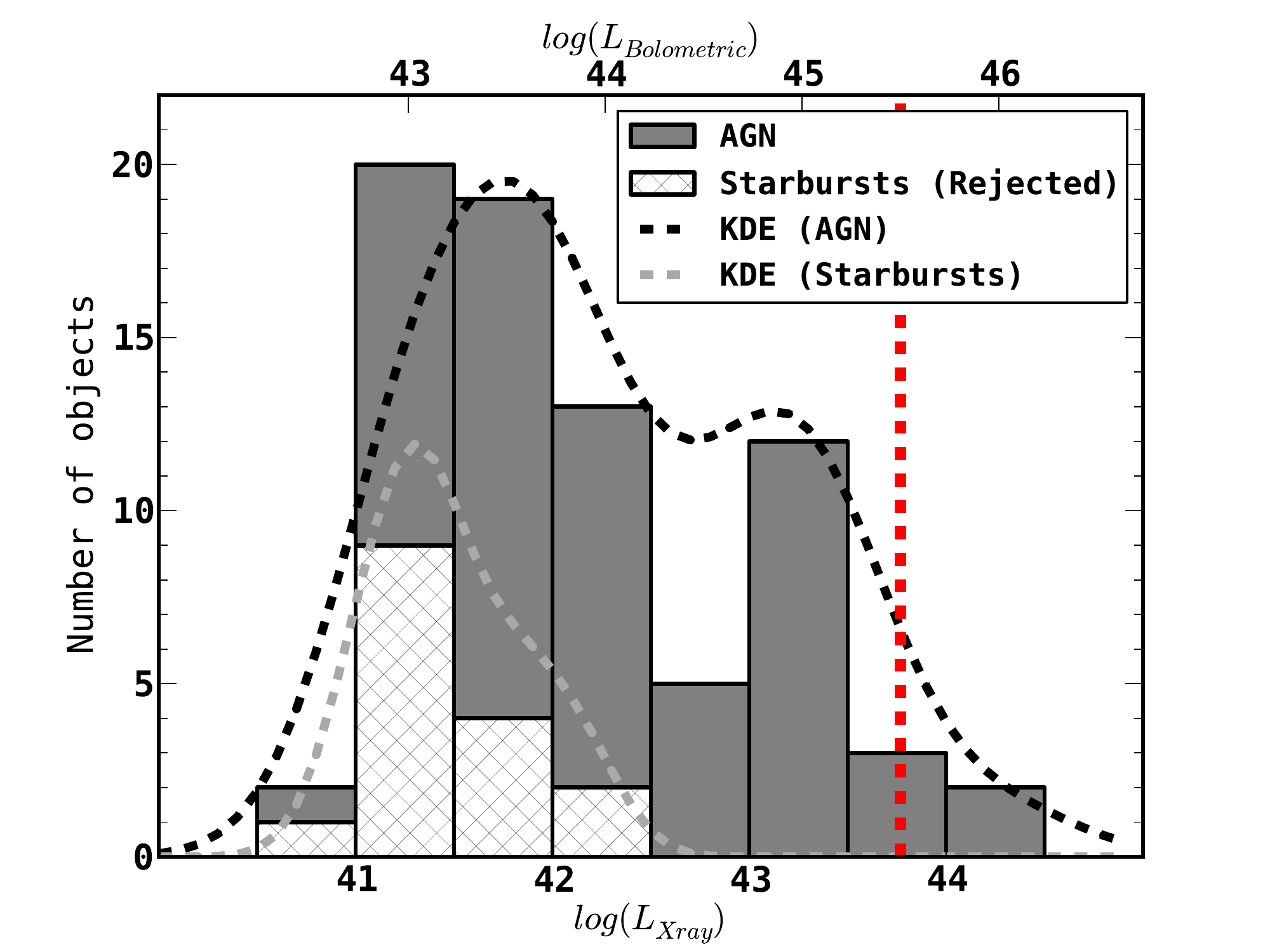}
\caption{Basic properties of the sample. \textit{Left panel:} Absorption corrected X-ray and bolometric luminosities of X-ray sources in the CDFS 4Ms Catalogue (grey dots), the sample used in this paper (red circles).  \textit{Right panel:} Histograms and kernel density estimators (KDE) of bolometric luminosities in the redshift range for all X-ray-sources used, objects rejected as starbursts are shown as cross-hatched. In both panels, the dotted red lines show the redshift range used, the dashed red line shows a bolometric luminosity of $\log(L_{bol} [erg/s])=45.5$. This is the luminosity above which mergers are thought to dominate AGN triggering \citep{somerville_semi-analytic_1999,somerville_semi-analytic_2008,hopkins_we_2013}.}
\label{F:sample}
\end{center}
\end{figure*}

The sample is selected from the Chandra Deep Field South (CDFS) 4 Ms data \citep{xue_chandra_2011}. X-ray selection provides a minimally biased sample over a wide range of luminosities. It is least affected by obscuration and at the great sensitivity of the data used, complete down to low AGN luminosities. It should however be noted that X-ray selection, even with the great depth of the 4Ms data will miss the most Compton-thick AGN. However, these systems are expected to be rare \citep[see e.g.][and references therein]{juneau_new_2011}. Figure \ref{F:sample} (left) shows the full 4Ms sample as well as the wavelength range chosen for this study. The chosen redshift range covers a maximum amount of dynamical range in X-ray luminosity while not covering too large a redshift range to have significant cosmological evolution within the sample ($\lesssim$ 2 Gyr in cosmic time). Surface brightness dimming effects are minimal. Throughout the paper, we use absorption corrected rest-frame 0.5-8keV luminosities in erg/s from \citet{xue_chandra_2011}. X-ray sources are matched to the H-band using a 1 arcsecond aperture.

From the 4Ms CDFS Sample, we study all 76 objects covered in CANDELS in the redshift range $0.5 < z < 0.8$. Additionally, we reject objects with soft X-ray spectra ($\Gamma >1$, where $\Gamma$ is the effective photon index) that lack a point source detection in the $F160W$ data since these sources are likely starbursts. We caution that extreme Compton thick sources can be potentially rejected using this method and this method is therefore not 100\% effective in identifying starbursts \citep{juneau_new_2011}. The rejection affects sources with detections in only a single Chandra band and very low luminosities (see Fig. \ref{F:sample}), leaving a sample of 60 AGN in the field. The X-ray luminosity distributions of the AGN as well as the rejected starburst sources are shown in Fig. \ref{F:sample} (right). Note that the redshift range chosen for this study includes a cluster of galaxies at a redshift of $\sim$0.75 \citep{salimbeni_comprehensive_2009,castellano_x-ray_2011}. The cluster has a $M_{200} \sim 3 \times 10^{14} M_{\odot}$ and a velocity dispersion of $\sigma \sim 630$km/s. 

\subsection{Matched Sample}
\label{S:match}

For the control sample, we use catalogues by \citet{dahlen_detailed_2010}, including photometric data over a wide range of wavelengths \citep[see]{giavalisco_great_2004,dahlen_detailed_2010}  as well as a compilation of spectroscopic redshifts \citep{cristiani_first_2000,croom_small-area_2001,dickinson_color-selected_2004,le_fevre_vimos_2004,stanway_three_2004,strolger_hubble_2004,szokoly_chandra_2004,van_der_wel_fundamental_2004,doherty_campanas_2005,mignoli_k20_2005,roche_deep_2006,ravikumar_new_2007,popesso_great_2009,vanzella_great_2008}.

\begin{figure}
\begin{center}
\includegraphics[width=10cm]{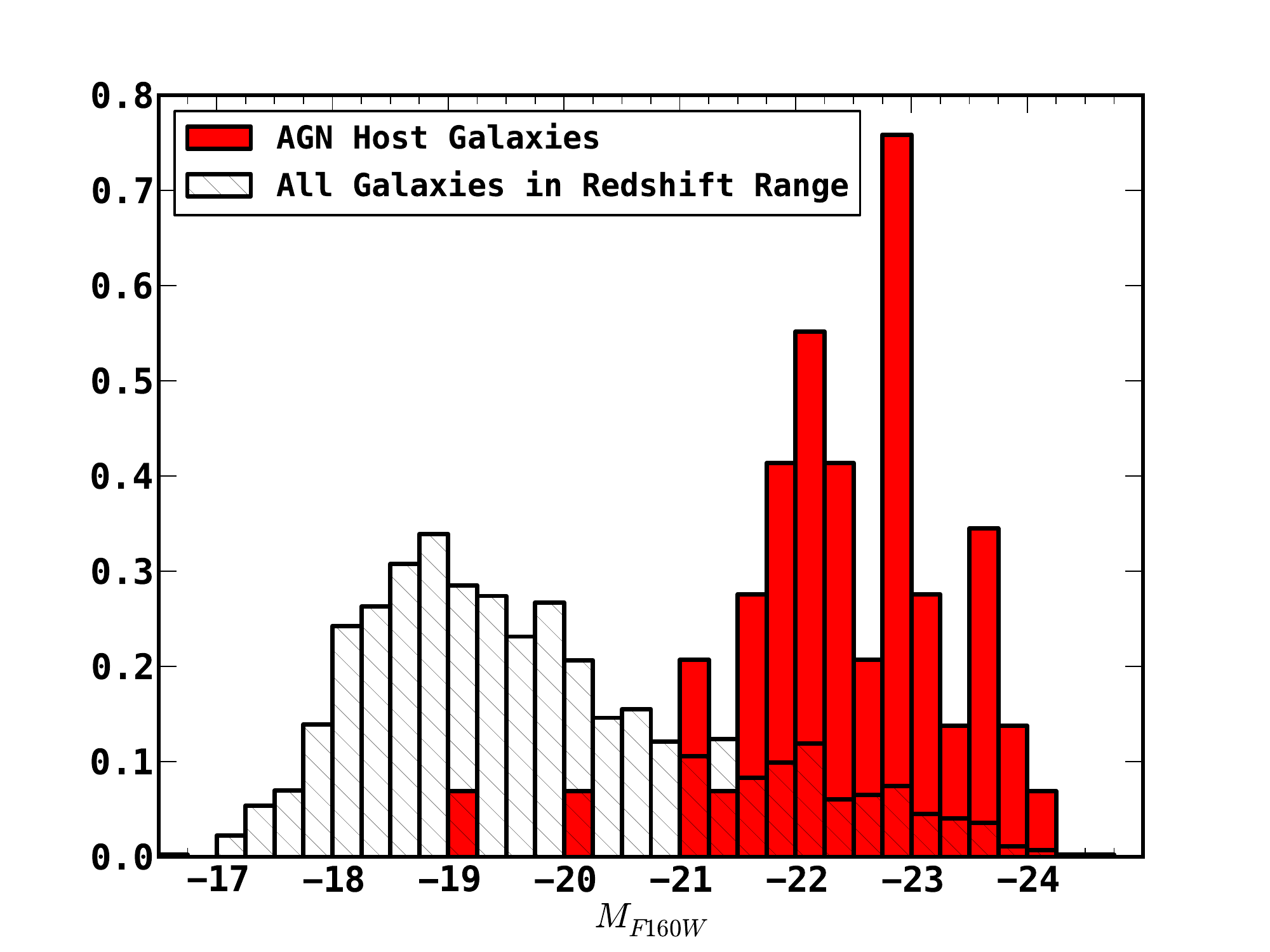}
\caption{Histograms showing the distributions of the AGN Host galaxies (red) as well as all galaxies in the redshift range of the study (hatched). Note that since the histograms are both normalized to integrate to one, this histogram does not represent the total number of galaxies in both samples.}
\label{F:galhist}
\end{center}
\end{figure}

\begin{figure}
\begin{center}
\includegraphics[width=10cm]{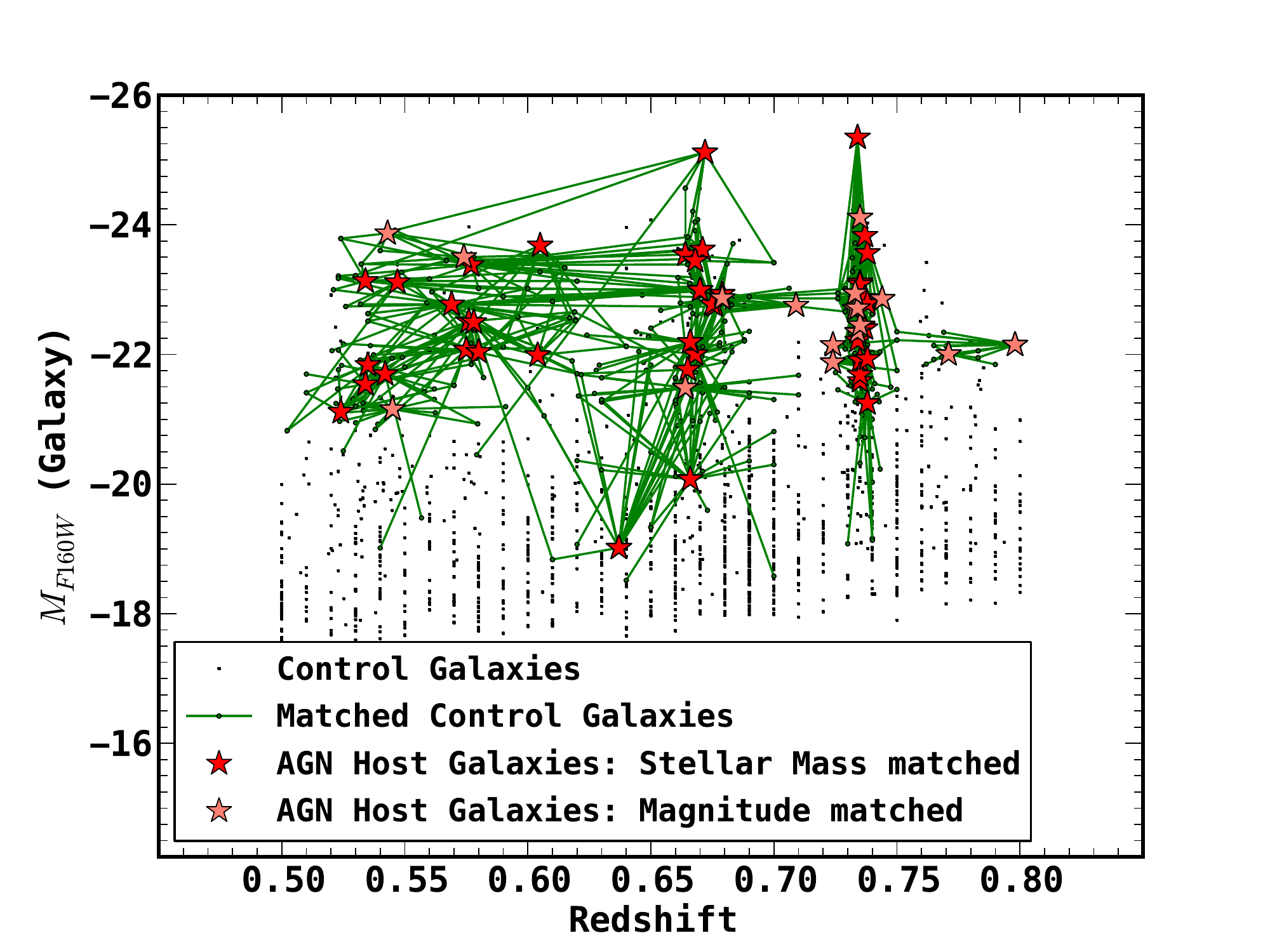}
\caption{Matching performed for all AGN, the black dots show the full control sample. The red stars mark the AGN, the green dots show galaxies that have been matched as control galaxies, the green lines connect them to the AGN they have been matched to. Note the cluster at a redshift of $z\sim0.75$.}
\label{F:match}
\end{center}
\end{figure}

The AGN host galaxy sample is dominated by massive galaxies, while the full galaxy sample in the same redshift range generally contains many more lower mass galaxies. This is mostly an effect of detection probability since low mass galaxies have lower mass black holes which makes detection at equal Eddington rate less likely \citep{aird_primus:_2011}. The histograms of AGN host galaxies and the parent control sample are shown in Fig. \ref{F:galhist}. From this parent control sample, we create a control sample by matching between 5 and 25 galaxies to each AGN (Fig. \ref{F:match}).

\begin{figure}
\begin{center}
\includegraphics[width=10cm]{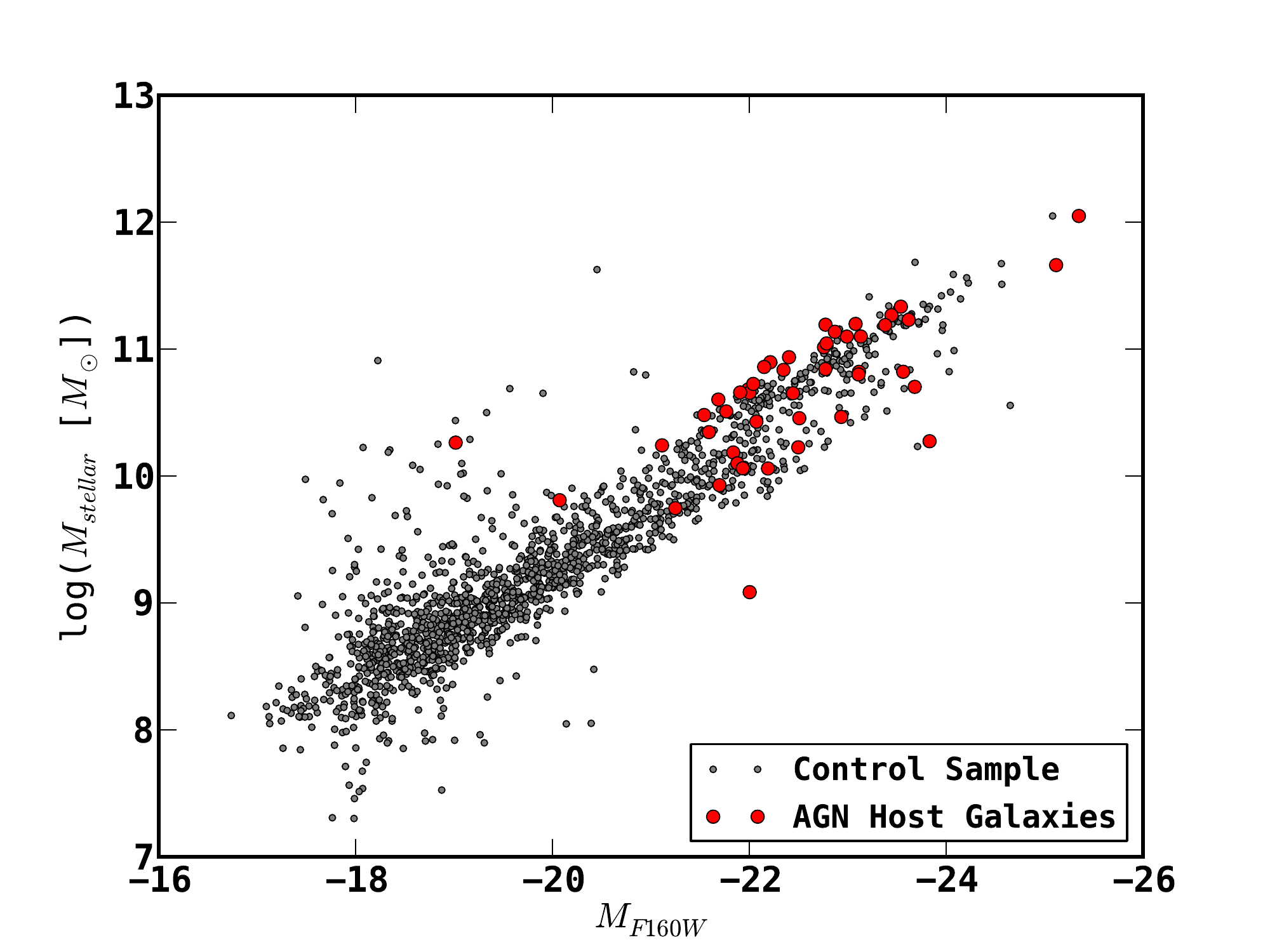}
\caption{Comparison between stellar masses and absolute H band magnitude for full control sample (grey dots, all galaxies in given redshift range) and AGN host galaxies (red filled circles).}
\label{F:stellarmass}
\end{center}
\end{figure}

Control galaxies are matched in absolute $F160W$ (host) galaxy magnitude and stellar mass from \citet{santini_enhanced_2012} as well as redshift. The stellar masses are derived by fitting the SEDs of the AGN with a mixture of galaxy and QSO templates, for a more detailed description of the process, we refer the reader to \citet{santini_star_2009} and \citet{santini_enhanced_2012}. The stellar mass is traced rather well by the H band magnitude (Fig. \ref{F:stellarmass}) with a scatter of $\sim$0.25 dex in stellar mass. Some of this scatter in stellar mass is explained by differences in star formation histories (i.e. starburst age, reddening).

Initial matching is performed by selecting galaxies with $\Delta z=0.05$ and $\Delta m=0.1$, stellar masses are matched within 10\%. If no sufficient number of matches are found, the criteria are relaxed. As can be seen in Figure \ref{F:match}, most galaxies are matched isotropically in redshift and magnitude. At lower galaxy magnitudes, AGN hosts can be matched to control galaxies with relatively large magnitude differences ($\sim0.5\Delta$mag). Additionally, two host galaxies are amongst the most luminous and massive galaxies in the field and redshift range (see Fig. \ref{F:match}). These AGN hosts are matched non-isotropically to control galaxies of lower mass. While this is not optimal, we simply lack appropriate optimally matched galaxies for these AGN.  However, we find that the morphological properties (in particular asymmetry and ellipticity) do not show a strong trend with absolute galaxy magnitude. Additionally, a large percentage of objects are closely matched. Our results hold when omitting the two AGN with the most anisotropic matching. The AGN with anisotropic matching do not belong to the bin with the most luminous AGN.

Objects for which stellar masses are not available are matched in H band galaxy magnitude instead (see Section \ref{S:fit} for a description of fits described to derive AGN host magnitudes). For AGN located in the cluster at $z\sim0.75$, we reject all control galaxies that are outside the clusters redshift range. Since the cluster environment can have a strong influence on galaxy morphological parameters \citep{dressler_galaxy_1980,adams_environmental_2012}, matching AGN hosts to control galaxies located in very different environments could introduce biases. The results of our study hold when matching is performed purely in H band magnitude.

\subsection{Host Galaxy Fits}
\label{S:fit}

Galaxy fits are performed using \textsc{Galfit} \citep{peng_detailed_2002}. Empirical PSFs were derived from stacking the images of several isolated and unsaturated stars in the field.  In order to provide a  more accurate description of the central region, we replaced the inner-most  pixels (within a radius of 3 pixels from the center) with a simulated PSF generated with the TinyTim package \citep{krist_simulation_1995}.  The TinyTim PSF was dithered and drizzled in the same manner as the observations, and normalized such that the total flux of  the newly constructed hybrid PSF model is the same as that of the stacked star. We found this hybrid PSF accurately reproduced the growth curves of stars out to 3".  Further details on the PSF models can be found in \citet{van_der_wel_structural_2012}.

For the AGN  host galaxies, we use a mixture of point source and Sersic component (with both ellipticity and radius left as a free parameter). If necessary, a second Sersic component is added. In most cases, a fit with a point source and single Sersic yielded a good fit with minimal residuals. The goodness of fit was judged both based on the $\chi^{2}$ values given by \textsc{Galfit} and visual inspections of the residuals. It was only necessary to add a second sersic component in four cases out of the 60 AGN studied. In these cases, the fits diverged for a single Sersic component or resulted in strong residuals indicative of the presence of a component not accounted for in the fits. Since we use the point-source subtracted images rather than the residual images for further analysis, details in the host galaxy fits themselves do not strongly affect our results.

The image resolution of 0.1" corresponds to about 0.5 kpc in the redshift range of our sample. A compact bulge or starburst might therefore be fit as a point source. For all AGN that require point source fits, we therefore check the colour of the central point source using ACS images. Colours of central point sources are blue, consistent with AGN. Starbursts however have similar colours. We find that contributions of central point sources in low-luminosity AGN are weak. There is the possibility that AGN contributions are overestimated at low X-ray luminosity, however, our results hold when omitting AGN with $\log(L_{X} < 42)$ and therefore this does not affect the overall conclusions of this study.

All Control galaxies are fit with a single Sersic component and no point source to determine their Sersic indices as well as ellipticities. To mimic residuals from PSF fitting performed for AGN hosts, we create fake AGN for the morphological analysis. Point sources with magnitudes matched to the corresponding AGN magnitude are added to its matched control galaxies. The 'fake AGN' are then fit with a point source and Sersic model. A more detailed discussion of the influences of residuals on morphological parameters can be found in Appendix \ref{S:appendix}.

\subsection{Quantitative Morphology Measures: Asymmetry}
\label{S:morphology}

For quantitative morphology measurements, we use Asymmetry A which is part of the CAS (\textbf{C}ompactness, \textbf{A}symmetry, \textbf{S}moothness) classification system \citep{conselice_relationship_2003}. Tests on our data showed that the concentration C was most sensitive to PSF residuals since it works by measuring how centrally concentrated galaxies are, in agreement with previous studies \citep{grogin_agn_2005}. Due to the limited resolution of our data, Smoothness S is not used. Asymmetry is found to trace major and minor mergers well \citep{lotz_effect_2010,lotz_effect_2010-1}. Additionally, we will use the ellipticity from the \textsc{galfit} host galaxy fits. Relaxed post-merger systems could show lower ellipticities than galaxies of similar mass that have not undergone a merger recently. A detailed description of the morphology measures and possible errors due to PSF residuals can be found in Appendix \ref{S:appendix} below. Here we provide a brief overview.

\begin{figure*}
\begin{center}
\includegraphics[width=18cm]{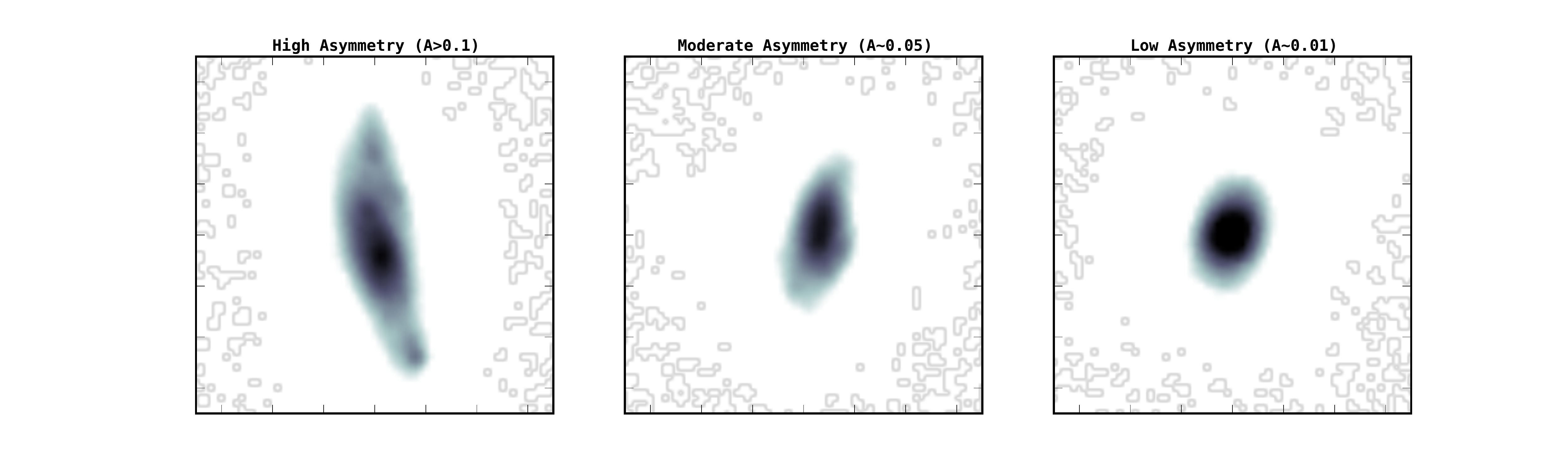}
\caption{Three example images showing different levels of asymmetry in three AGN host galaxies. All images are on a logarithmic scale and 4.2"$\times$4.2" in size.}
\label{F:examples}
\end{center}
\end{figure*}

Asymmetry A is defined as:

\begin{equation}
A \equiv \sqrt{ \dfrac{ \sum \frac{1}{2} \times (I_{0} - I_{180})^{2} }{\sum I_{0} ^{2}}  }
\end{equation}
where $I_{0}$ is the flux in each pixel and $I_{180}$ is the flux in each pixel rotated by 180$^{\deg}$ \citep{conselice_relationship_2003}. Different from \citet{conselice_asymmetry_2000} and \citet{conselice_relationship_2003}, we do not subtract background asymmetry. However, as shown in \citet{conselice_asymmetry_2000}, this will have little effect for the typically bright galaxies used in this study. For the purpose of this study, we use Sextractor segmentation maps to avoid including noise from the background into the measurement. These maps determine the region over which the galaxy is detected. In the following, we will discuss different influences on the measured asymmetry. Centring is performed following \citet{conselice_asymmetry_2000}. We have ensured that the algorithm generally reaches a well-defined minimum. Visual inspection of all AGN hosts and a randomly chosen subset of the control sample is performed to make sure the central point determined by the algorithm determines the center of the galaxy correctly. Three galaxies with different levels of asymmetry are shown in Fig. \ref{F:examples} as a reference for the reader.

Central pixels are in some cases affected by PSF residuals. While simulating 'fake AGN' ensures that this is also the case for the control sample, we still do not wish these pixels to dominate the overall asymmetry measurement. Hence, the central area of all objects is masked using a circular aperture with a radius of two pixels. The exact size of the mask does not change the overall result and visual inspections show that for the AGN magnitudes common in our sample, these aperture sizes cover the corrupted pixels while not masking uncontaminated areas of the host galaxy. More details can be found in Appendix \ref{S:appendix}.

In order to keep asymmetry measures comparable independent of optical AGN magnitude, we use the same mask size for all objects, independent of AGN magnitude. While this also masks the central areas of galaxies not affected by PSF residuals, it ensures that the same physical areas are used for the calculation of the asymmetry for all sources. Ideally, the mask size would be adjusted to the size of the galaxy. However, since AGN hosts and control are carefully matched, this should not result in a biases measurement.

\subsection{Human classifiers}
\label{S:humans}

We use results from human classifiers for comparison and consistency checks. The galaxies were classified by CANDELS team members in the $H/F160W$ band. Bluer bands are also inspected to facilitate the classification \citep[][, Kartaltepe et al. in prep]{kartaltepe_goods-herschel_2012}. The boundaries of the galaxies are defined used \textsc{Sextractor} segmasks. In a first step, the classifier is asked to decide between four main morphological classes: spheroid, disk, irregular/peculiar and compact/unresolved. Additionally, the human classifier is asked to decide if the galaxy meets any of the following interaction classes: merger (a clearly interacting system with disturbances), interaction (interaction of two distinct objects within or beyond the segmentation mask) or non-interaction companion (galaxy has a companion that is not clearly disturbed). This classification scheme is the same as used in \citet{kocevski_candels:_2012}.

\subsection{Note on confidence intervals used}
\label{S:statistics}

Binomial probabilities and their confidence intervals for the AGN sample are derived using a beta statistic \citep{cameron_estimation_2011}. As stated in Section \ref{S:match}, several control galaxies are matched to each AGN to increase statistical power. Due to the fact that the number of matches differs between AGN, confidence intervals for the control sample cannot be derived straightforwardly using the beta statistics since it does not account for weighting. We therefore use a jackknife method in which we randomly choose a subset of five matched control galaxies for each AGN and then derive the resulting value for the binomial probability. This random matching is repeated 100 times and the confidence intervals are derived from the final distribution. Jackknife methods are also used to determine the expectation values and error in distribution moments for the control sample.

\section{Results: Morphologies of AGN Hosts}
\label{S:results}

We now compare the morphological properties (Sersic indices, ellipticities and asymmetries) of the AGN host galaxies to those of the control galaxies. The aim is to determine if the data are consistent with the null hypothesis that AGN hosts are drawn randomly from the sample of their control galaxies. Note that when plotting X-ray luminosity, we plot the X-ray luminosity of the matched AGN for the control galaxies.

\begin{figure*}
\begin{center}
\includegraphics[width=8.5cm]{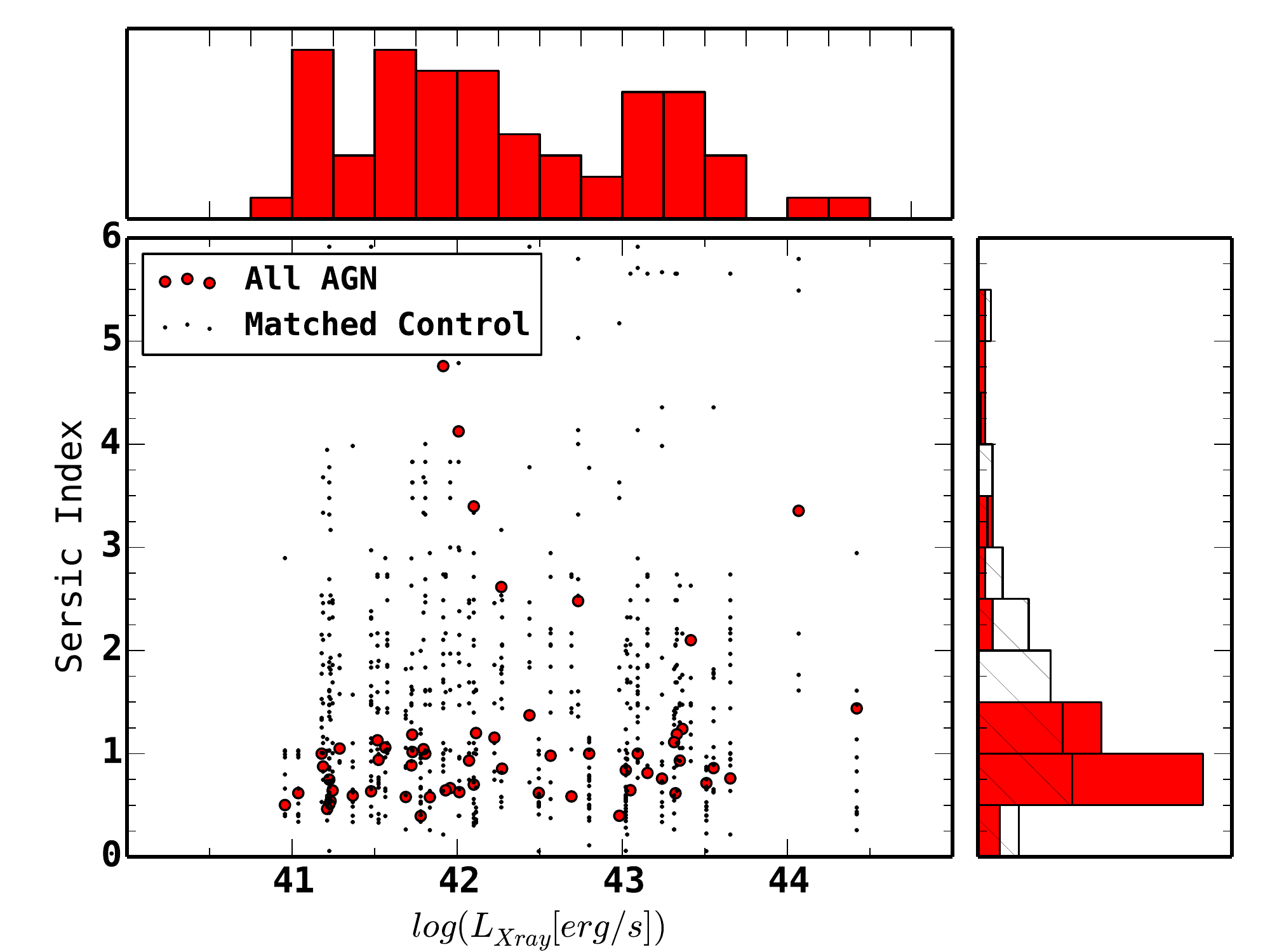}
\includegraphics[width=8.5cm]{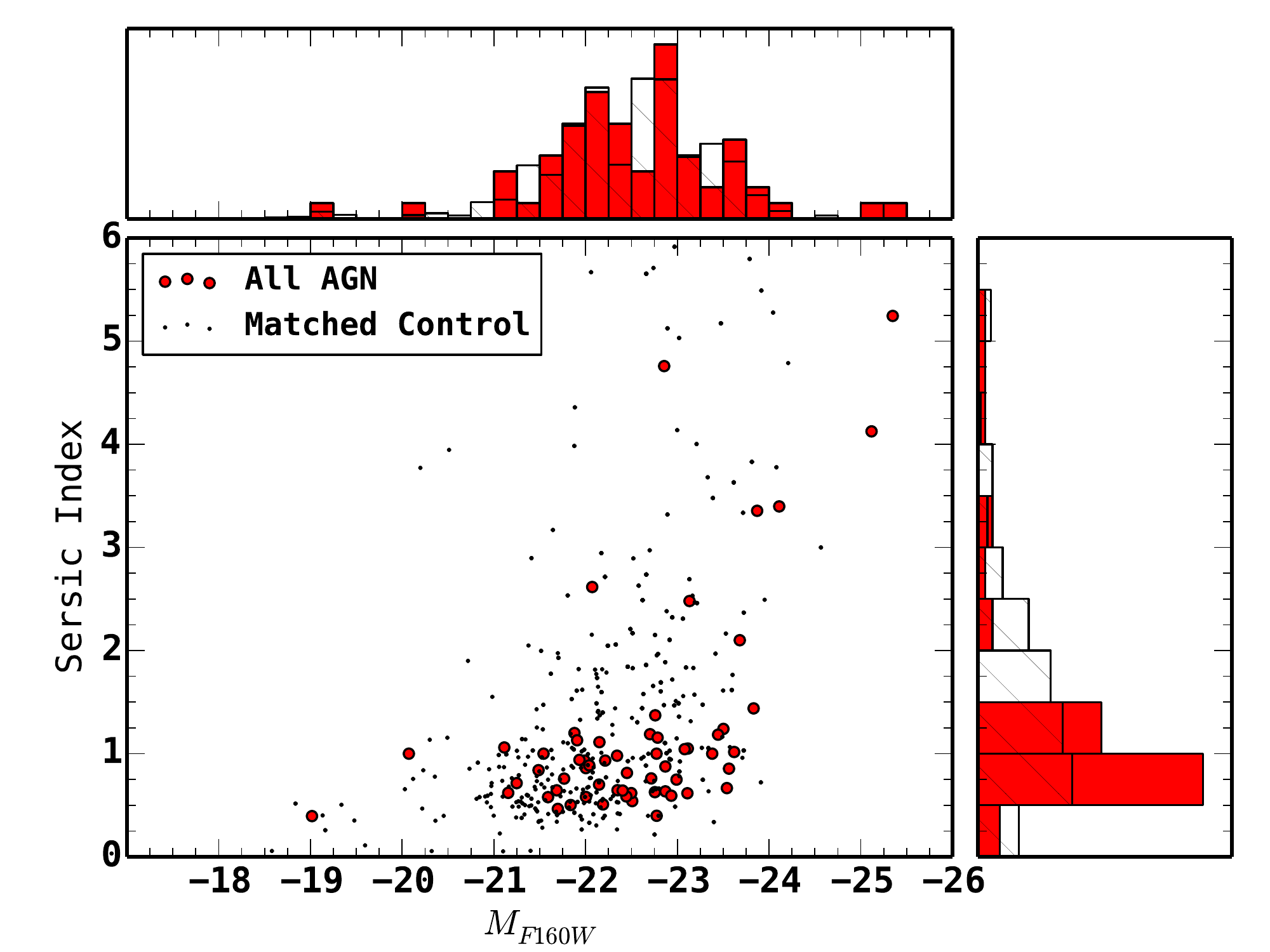}
\caption{Sersic index as a function of AGN Luminosity (left) and AGN Host magnitude (right) for AGN (red circles) and control (grey dots). For control galaxies, the X-ray luminosity of the matched AGN is plotted. Projected histograms are shown for the AGN (red) and control sample (hashed). There are no statistically significant differences between the two samples.}
\label{F:sersic}
\end{center}
\end{figure*}

We compare the distribution of Sersic indices for the AGN and control sample (Fig. \ref{F:sersic}). Two sample tests show no statistically significant difference between AGN host galaxies and the control sample. As expected, Sersic indices are higher for more luminous host galaxies, there is however no strong trend with X-ray luminosity. 

\begin{figure*}
\begin{center}
\includegraphics[width=8.5cm]{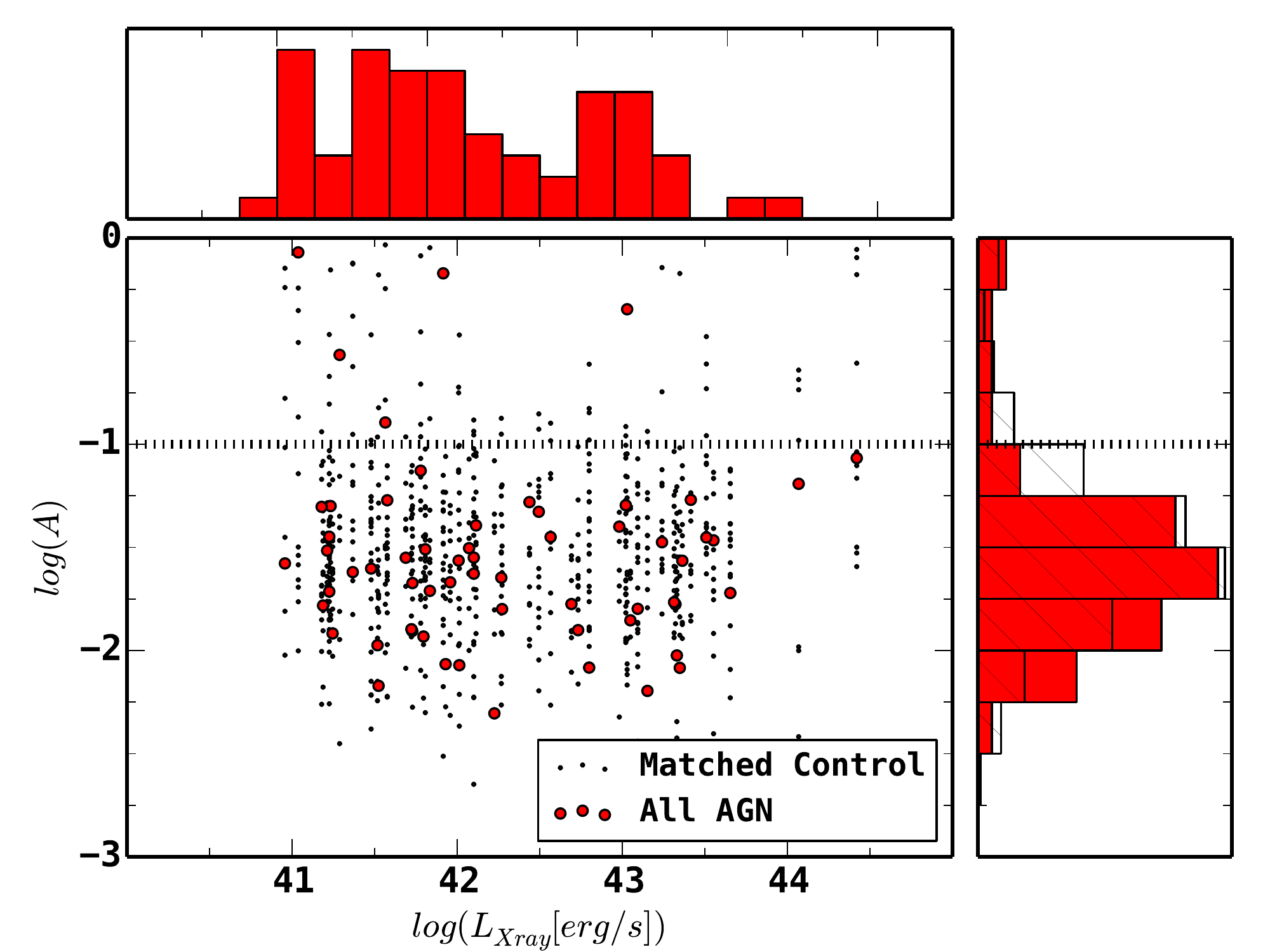}
\includegraphics[width=8.5cm]{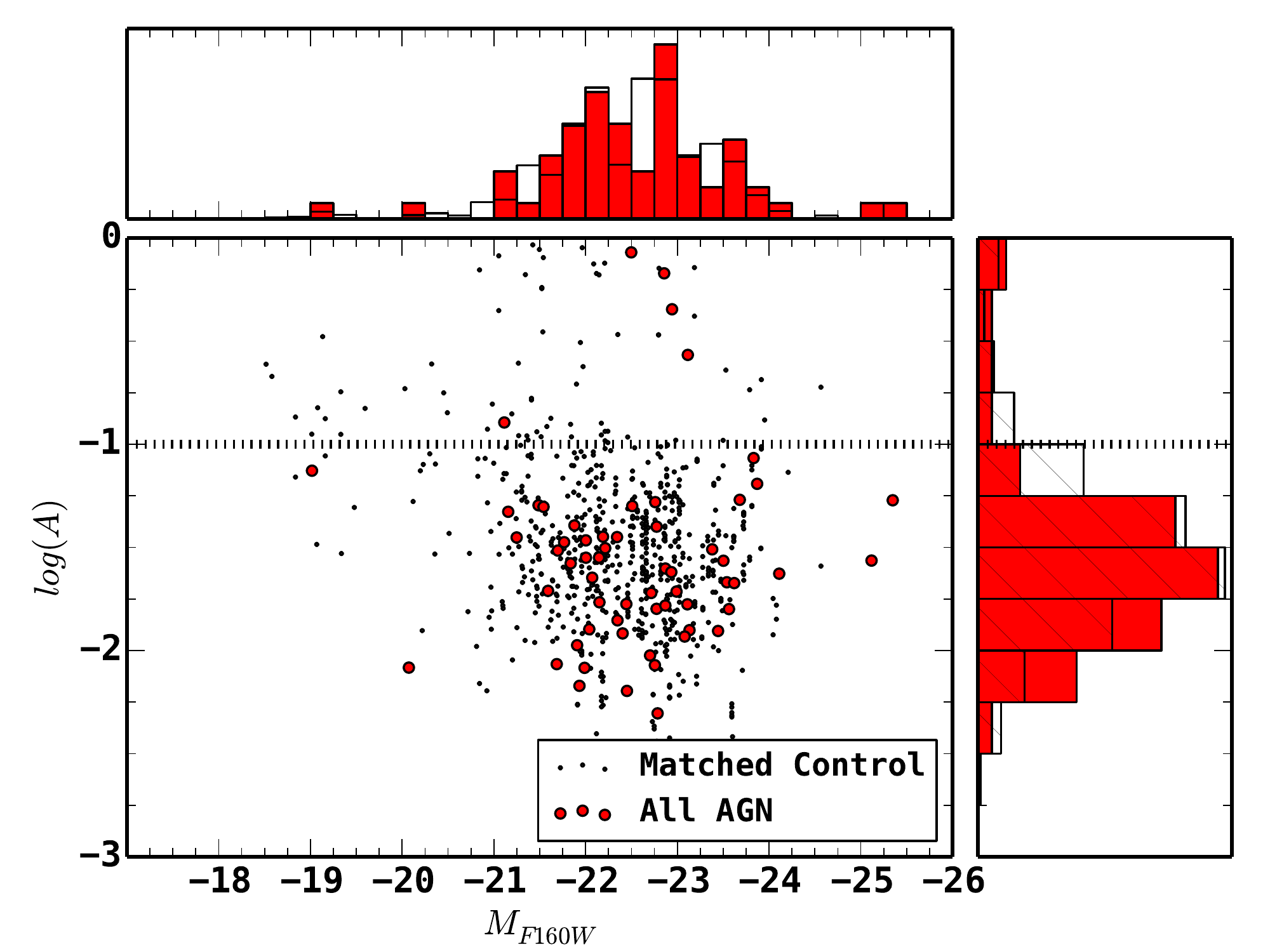}
\caption{Asymmetry as a function of Xray luminosity (left) and AGN host galaxy magnitude (right) for AGN (red circles) and control (grey dots). For control galaxies, the X-ray luminosity of the matched AGN is plotted. Projected histograms are shown for the AGN (red) and control sample (hashed). There are no statistically significant differences between the two samples.}
\label{F:asym}
\end{center}
\end{figure*}

Next, we analyse differences between the asymmetries of AGN host and control galaxies (Fig. \ref{F:asym}). The overall distribution of the asymmetries for the two samples show no significant differences. We also find no trend with absolute galaxy magnitude. We perform two- sample Kolmogorov-Smirnov (KS) and Mann-Whitney U (MW) tests between the AGN and control galaxies. These tests are calculated for both the full sample and sub-samples binned in both X-ray luminosity and absolute galaxy magnitude. The results are shown in Table \ref{T:stats_p}. The two-sample Kolmogorov-Smirnov test yields p=0.38 and the Mann-Whitney U test yields p=0.09. We hence fail to reject the null hypothesis that the two samples are drawn from the same parent population. When binned in X-ray and galaxy luminosity, the Kolmogorov-Smirnov and Mann-Whitney U test find no significant differences between the samples. We only find $p\leq0.05$ in a single bin. However, given the multiple tests (28) performed, this is consistent with the expected number of false positives (1.4).

While the statistical tests reveal no significant difference between the asymmetries, we do note that by eye the distribution of asymmetries appears to be slightly more skewed with a larger high asymmetry tail in the AGN host galaxies when compared to control. To determine if these differences are quantifiable, we calculate the moments of the distributions of asymmetries for the AGN and control galaxies. A jackknife method is used to determine the typical scatter in the moments derived for the control sample. We calculate the first four moments (mean, standard deviation $\sigma$, skew $\gamma$ and kurtosis $\kappa$). For the full sample, we find that both the skew and kurtosis are higher for the AGN sample compared to control. Both lie above the third quartile of the control sample distribution. This implies that the distribution for the AGN hosts has a tail towards \textit{larger} values of the asymmetry A and a \textit{higher} peak with more power in the tails. However, when dividing the sample into sub-bins in either AGN or host galaxy luminosity, the sample sizes are too small to determine if there is a trend in the moments. The full calculated properties are listed in Appendix \ref{S:appendix_stats}.

\begin{figure}
\begin{center}
\includegraphics[width=8.5cm]{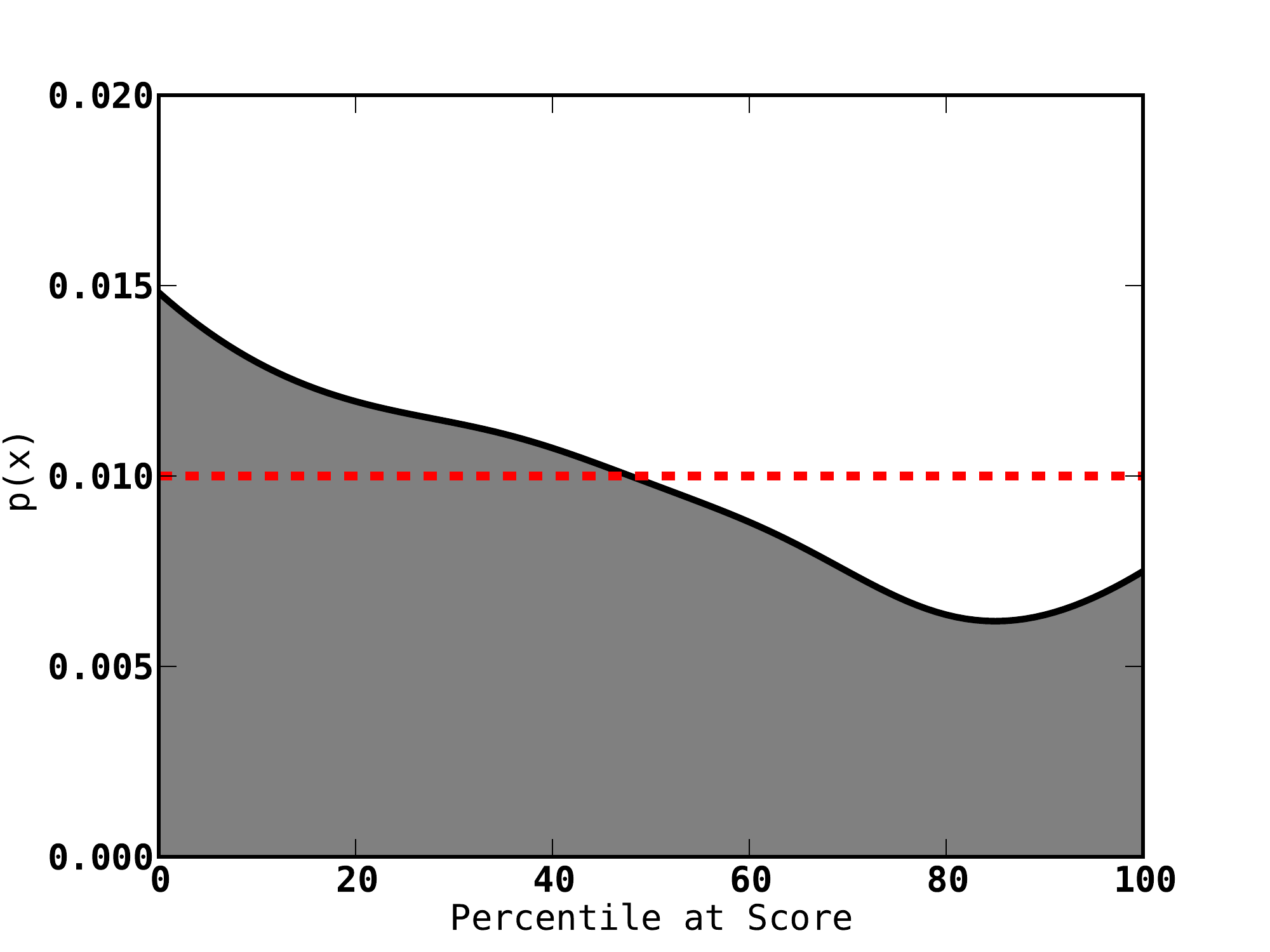}
\caption{Percentile distribution of AGN asymmetries with respect to their control sample. For each AGN, we calculate the percentile at score with respect to its matched control galaxies. The KDE is reflected off the boundaries. If the AGN host asymmetries were drawn randomly from the same distribution as the matched control galaxies asymmetries, the distribution should be flat (indicated by the dashed red line). The difference is however of low statistical significance (p=0.098).}
\label{F:percentile}
\end{center}
\end{figure}

Another way to compare the asymmetries of AGN hosts and control galaxies is to compare the values for each single AGN to its sample of control galaxies. We therefore analyse the percentiles of scores of each AGN host galaxy asymmetry with respect to its matched control galaxies. For each AGN, we calculate the cumulative density function of the asymmetries of its control sample. From this distribution, we then calculate the percentile at score for each AGN. For example, an AGN with an asymmetry equal to the median asymmetry of its control sample will have a percentile at score of 50\%, while an AGN having asymmetry higher than all its matched control galaxies will have a percentile at score of 100\%. If the AGN hosts were drawn randomly from the sample of its matched hosts, the distribution of percentiles at score for the full AGN sample should be flat since each AGN is equally likely to be sampled at each percentile at score. We show the kernel density estimator (KDE) of the percentiles at score for the full sample in Fig. \ref{F:percentile}. The distribution is not flat, showing excess at low percentiles at score as well as a small excess at very high percentiles at score. This is consistent with the eye ball assessment that the asymmetries appear on average somewhat lower in the AGN compared to control while showing a slight high-asymmetry excess. However, due to small number statistics, the differences are not statistically significant (p=0.098). 

\begin{figure}
\begin{center}
\includegraphics[width=8.5cm]{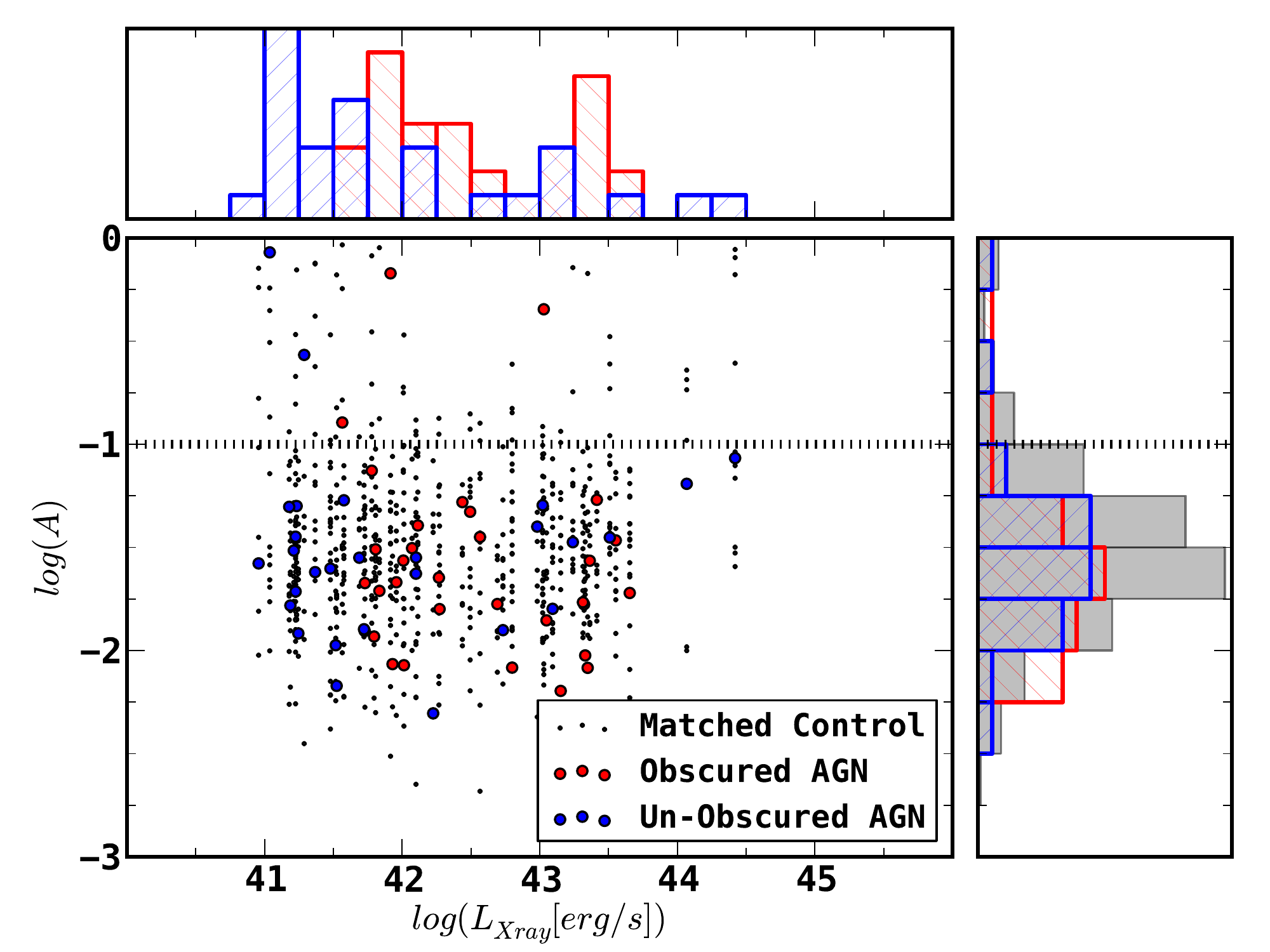}
\caption{Asymmetry as a function of x-ray luminosity for AGN (red and blue circles) and control (grey dots). The plot shows a comparison between x-ray obscured (red) and unobscured (blue) sources. For control galaxies, the X-ray luminosity of the matched AGN is plotted. Projected histograms are shown for the obscured AGN (red), un-obscured AGN (blue) and control sample (grey). There are no statistically significant differences between the samples.}
\label{F:asym_obs}
\end{center}
\end{figure}

An additional factor in asymmetry levels might be obscuration. While it is widely acknowledged that obscuration in AGN in the local Universe is mostly due to a dusty torus and therefore primarily a function of AGN orientation \citep{antonucci_unified_1993,urry_unified_1995}, some very young AGN might be in an earlier buried phase in which the obscuration is due to dust in the host galaxy and not the torus. We note that obscuration in the optical and the X-ray is not necessarily tracing the same obscuring material. Such a sub-sample of young obscured AGN might be more closely connected to mergers. The asymmetry distributions of AGN with X-ray effective photon indices $\Gamma < 1$ (X-ray obscured) and $\Gamma > 1$ (X-ray unobscured) are compared in Fig. \ref{F:asym_obs}. There are no differences between the two AGN sub-sample asymmetries (Table \ref{T:stats_p}), indicating that higher levels of obscuration do not lead to comparatively larger asymmetries.

\begin{figure*}
\begin{center}
\includegraphics[width=8.5cm]{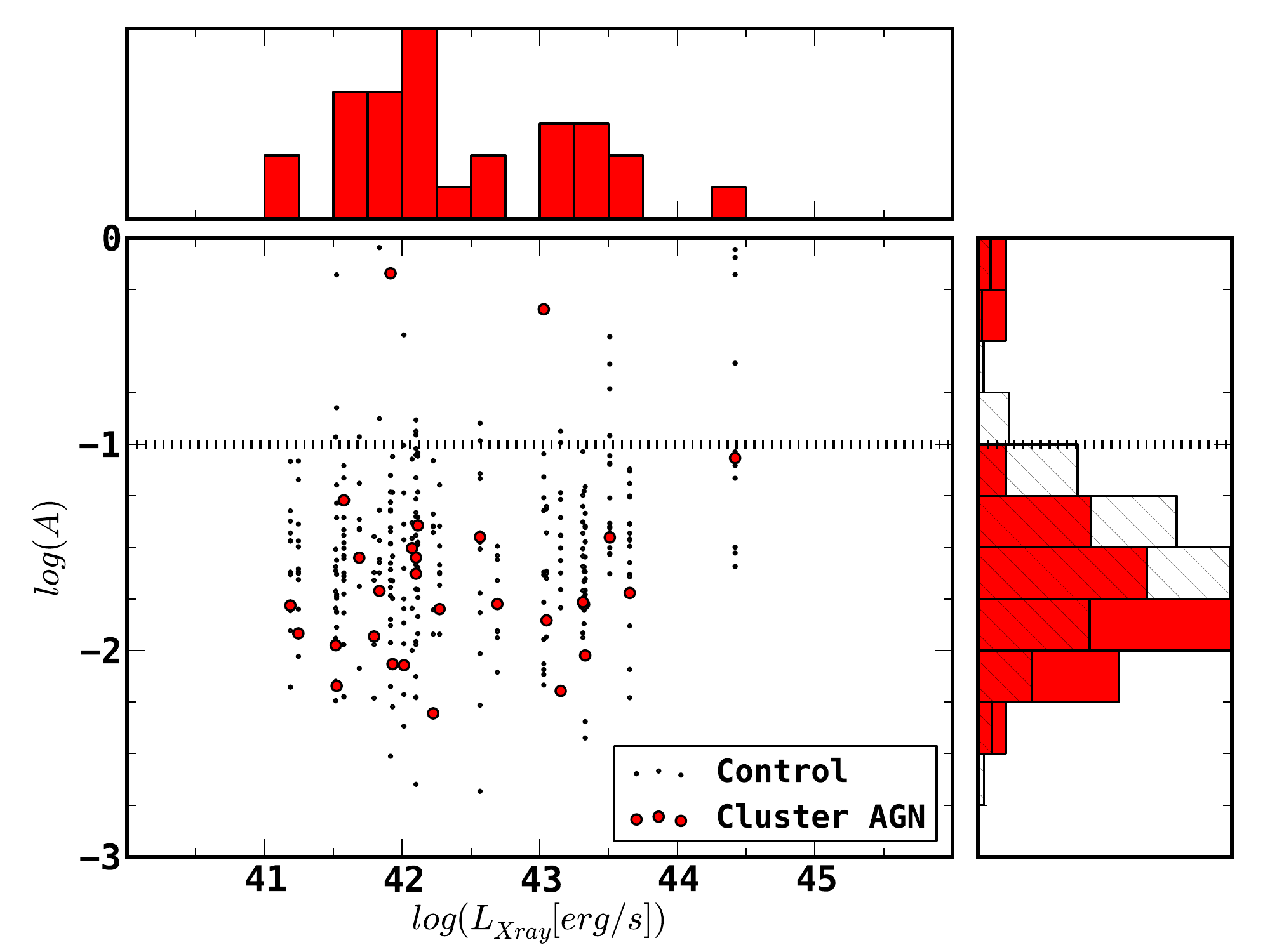}
\includegraphics[width=8.5cm]{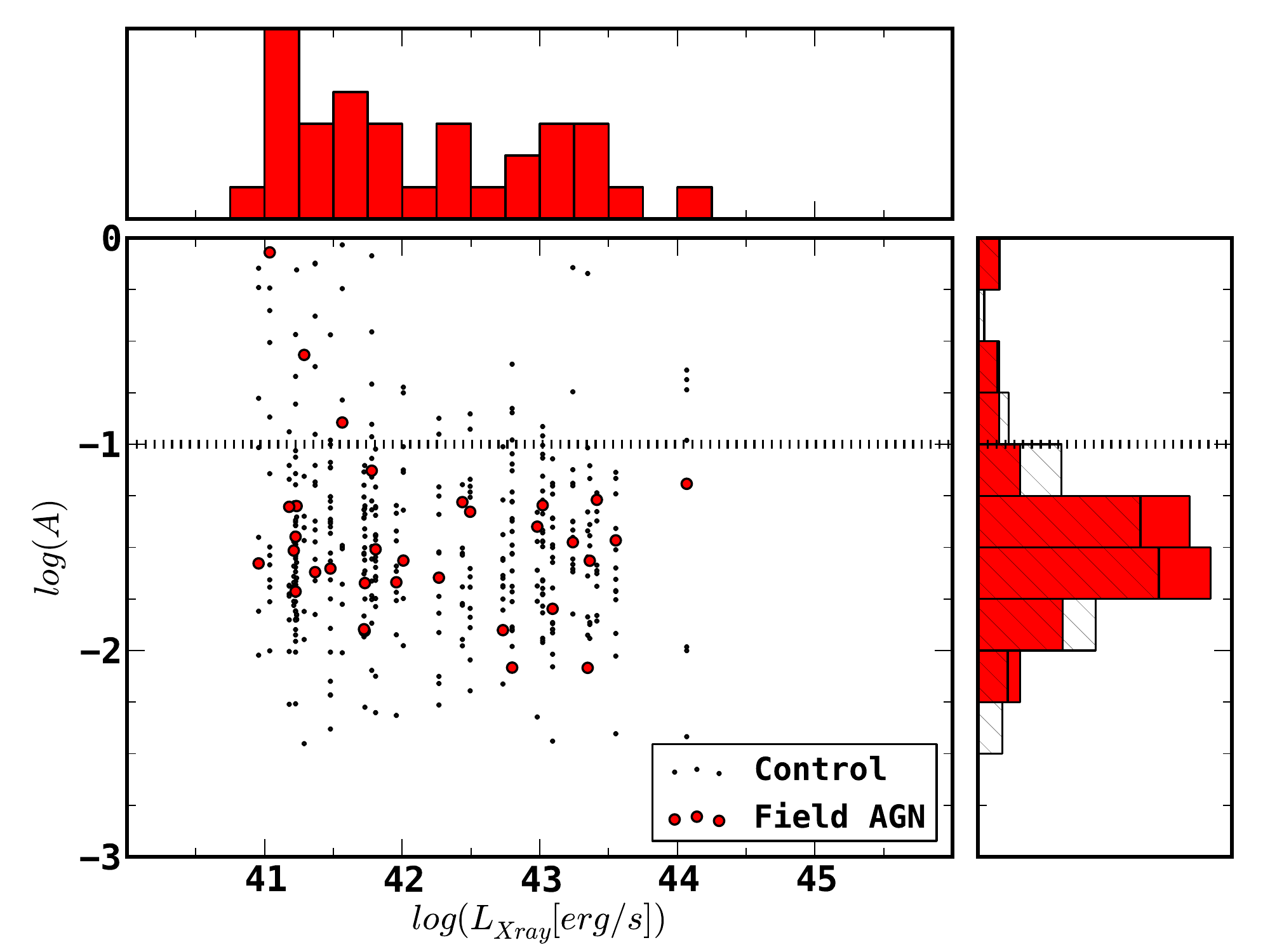}
\caption{Asymmetry in the cluster (left) and field (right) as a function of Xray luminosity for AGN (red circles) and control (grey dots), showing differences between the cluster and field. For control galaxies, the X-ray luminosity of the matched AGN is plotted. Projected histograms are shown for the AGN (red) and control sample (hashed). There are no statistically significant differences between the two samples in the field, however, in the cluster, the difference between AGN hosts and control is statistically significant (p$<$0.01).}
\label{F:asym_cluster}
\end{center}
\end{figure*}

As mentioned earlier, the redshift range studied contains a cluster of galaxies at z$\approx$0.75 \citep{salimbeni_comprehensive_2009,castellano_x-ray_2011}. We separately compare the host galaxies and their matched control galaxies in both the cluster and field (Fig. \ref{F:asym_cluster}). We find that - as for the full sample - the AGN hosts in the field are consistent with being drawn randomly from the control galaxies. However, in the cluster, we do find a statistically significant difference between the AGN and control galaxies (P $<$ 0.01) with AGN hosts having \textit{lower} mean asymmetries but \textit{higher} skew in the asymmetry distribution when compared to control galaxies.

\begin{figure*}
\begin{center}
\includegraphics[width=8.5cm]{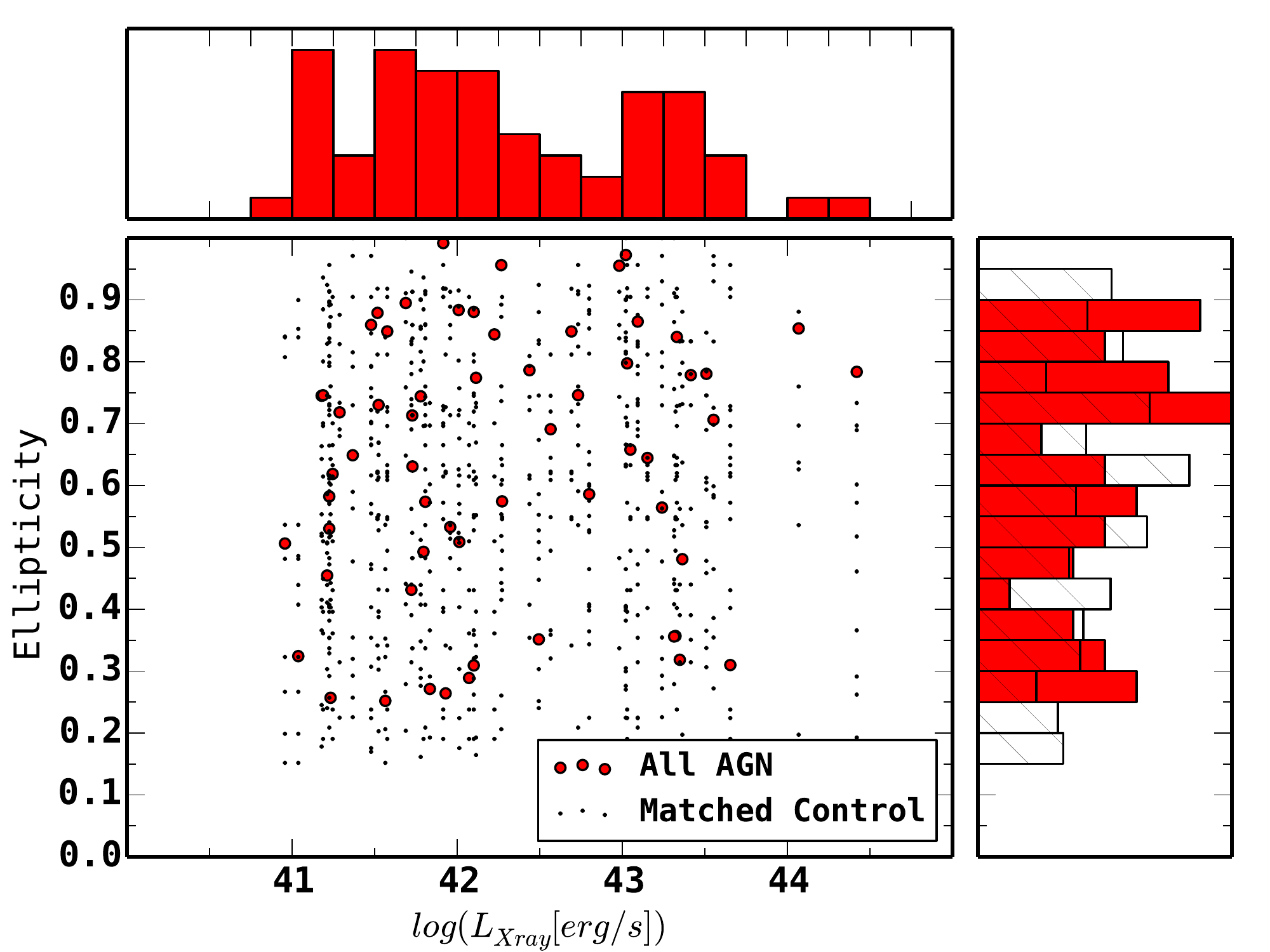}
\includegraphics[width=8.5cm]{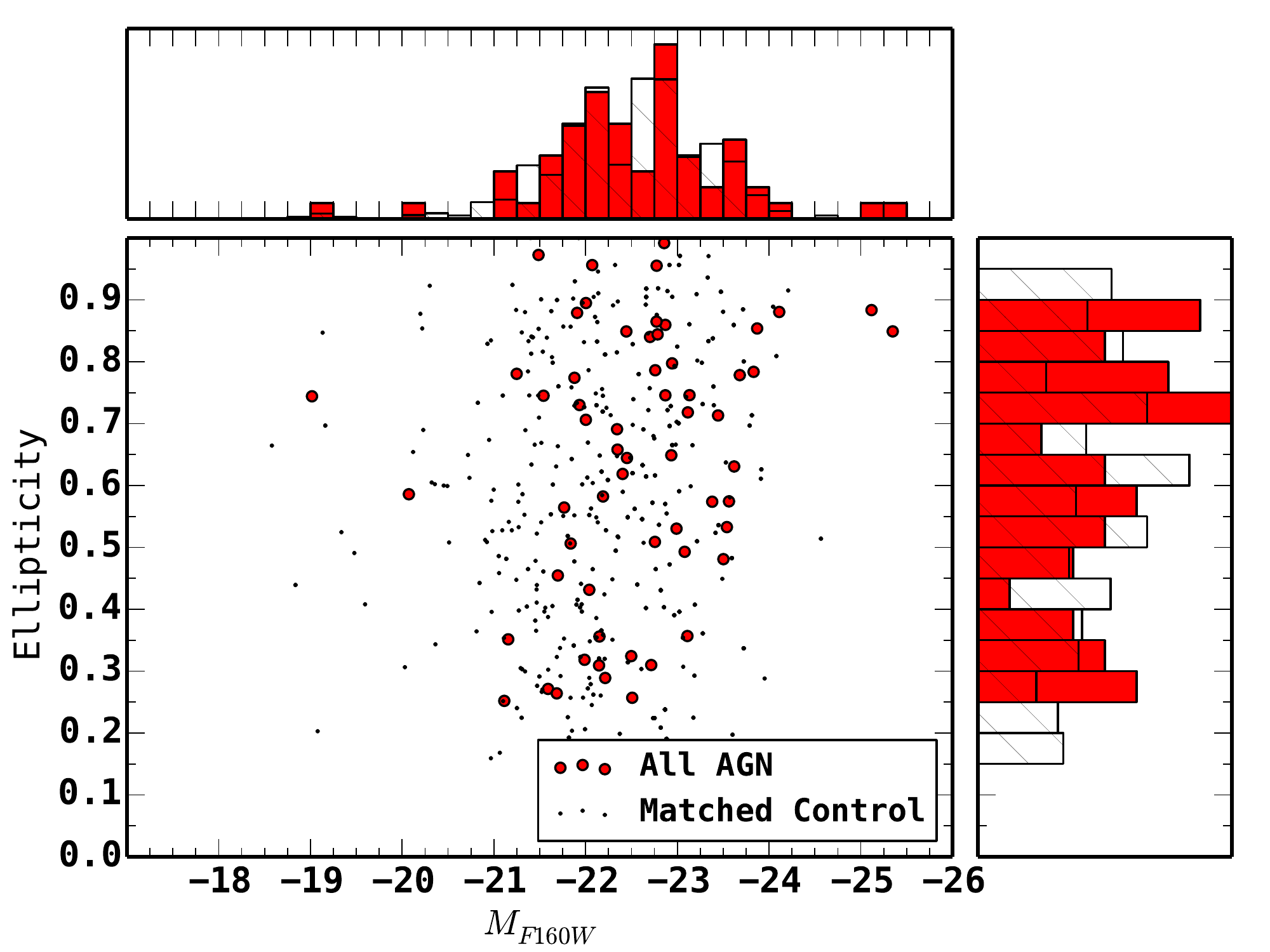}
\caption{Ellipticity as a function of AGN Luminosity (left) and AGN Host magnitude (right) for AGN (red circles) and control (grey dots). For control galaxies, the X-ray luminosity of the matched AGN is plotted. Projected histograms are shown for the AGN (red) and control sample (hashed). There are no statistically significant differences between the two samples.}
\label{F:ellipticity}
\end{center}
\end{figure*}

While asymmetry traces levels of disturbance in the host galaxy, more relaxed mergers will not be identified by this index (for example mergers between high ellipticity disk galaxies will generally reduce the ellipticity in the merged system). We thus additionally compare the ellipticities from the \textsc{galfit} galaxy fits between AGN hosts and matched controls. The results are shown in Fig. \ref{F:ellipticity}. There are no statistically significant differences between AGN host galaxies and control galaxies.

\begin{table}
\begin{minipage}{80mm}
\caption{Results from 2 sample Kolmogorov-Smirnov and Mann-Whitney-U tests comparing asymmetries of AGN hosts and matched control samples. Measure: morphological measure used for comparison; second columns: AGN property used for binning; p(KS): p-value for Kolmogorov-Smirnov tests. p(MW): p-value for Mann-Whitney-U test.}
\begin{tabular}{cccc}
\hline
Bin Property  & Bin Mean & p (KS) & p (MW)\\
\hline
$\log(L_{Xray})$ (All) & 42.27 & 0.34 & 0.09 \\ 
$\log(L_{Xray})$ & 41.18 & 0.85 & 0.21 \\ 
$\log(L_{Xray})$ & 41.59 & 0.18 & 0.07 \\ 
$\log(L_{Xray})$ & 41.91 & 0.56 & 0.27 \\ 
$\log(L_{Xray})$ & 42.33 & 0.66 & 0.40 \\ 
$\log(L_{Xray})$ & 43.04 & 0.10 & 0.12 \\ 
$\log(L_{Xray})$ & 43.60 & 0.85 & 0.25 \\ 
&&&\\
$M_{Host}$ (All) & -22.53 & 0.38 & 0.09 \\ 
$M_{Host}$ & -24.02 & 0.76 & 0.43 \\ 
$M_{Host}$ & -23.10 & 0.37 & 0.23 \\ 
$M_{Host}$ & -22.75 & 0.85 & 0.38 \\ 
$M_{Host}$ & -22.32 & 0.37 & 0.18 \\ 
$M_{Host}$ & -21.94 & 0.18 & \textbf{0.05} \\ 
$M_{Host}$ & -21.06 & 0.85 & 0.43 \\ 
&&&\\
All Cluster & -- & \textbf{0.0049} & \textbf{0.0098} \\
All Field & -- & 0.66 & 0.41 \\ 
&&&\\
Obscured vs Unobscured & -- & 0.31 & 0.17 \\ 
\hline
\end{tabular}
\label{T:stats_p}
\end{minipage}
\end{table}

\subsection{Comparison between quantitative measures and human classifiers}
\label{S:results_humans}

\begin{figure*}
\begin{center}
\includegraphics[width=8.5cm]{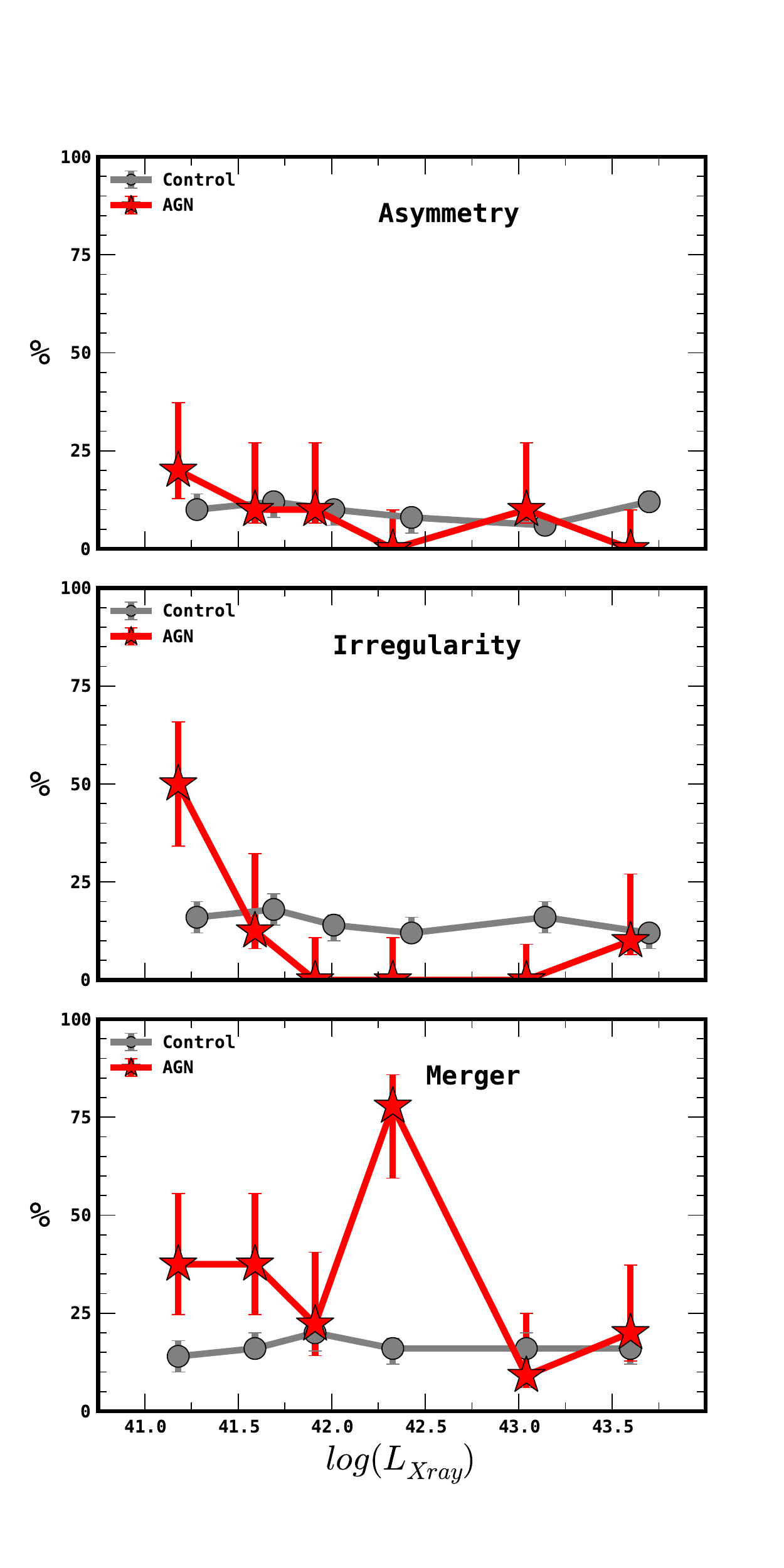}
\includegraphics[width=8.5cm]{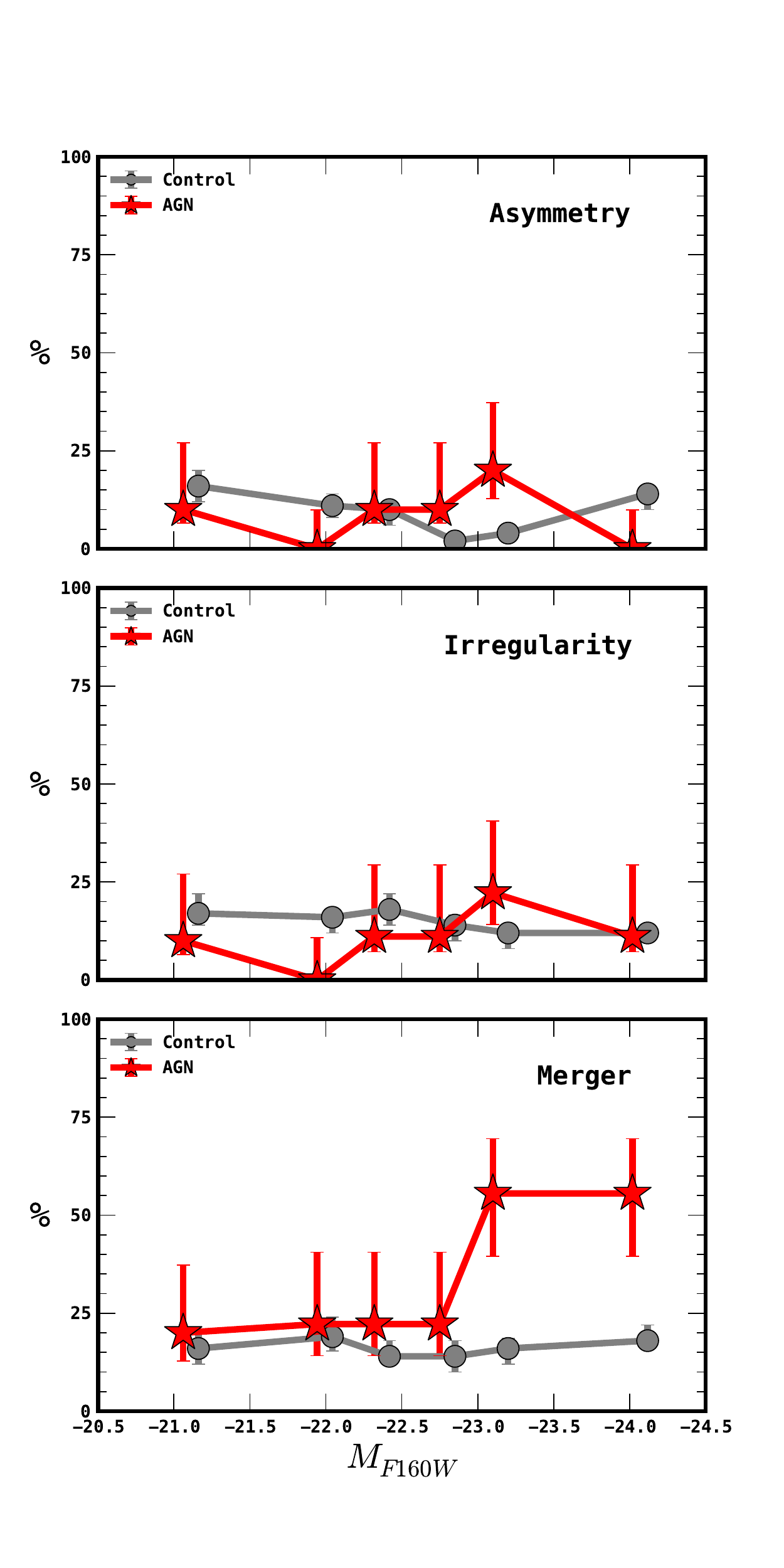}
\caption{Rates of disturbances in AGN host galaxies compared to control. The  \textbf{left} panel show the rates as a function of X-ray luminosity, \textbf{right} panel as a function of host galaxy absolute magnitude. The top panel is from quantitative asymmetry measurements, the 2 bottom panel rows are from human classifiers, in particular, we show irregularity (all galaxies showing some irregular features) and mergers (galaxies either showing clear signs of interaction or having close-by neighbours). Over-plotted are 68.75\% (1 $\sigma$) confidence intervals.}
\label{F:disrates}
\end{center}
\end{figure*}

In addition to the distributions of asymmetries, we also compare the probabilities of objects having high (A$>$0.1) asymmetries between AGN host galaxies and control samples. The cut-off A$>$0.1 is somewhat arbitrary, but visual inspection shows this to be a reasonable value above which all galaxies show clear disturbance (see also Fig. \ref{F:examples}). These rates are compared to different measures from the human classifier results described in Section \ref{S:humans} in Figure \ref{F:disrates}. 

In particular, we compare two different classifications, irregularity and merger. Irregularity encompasses all objects showing some disturbance or irregularity, even if they show a well-pronounced disk or spheroid component. The classifiers are asked to classify objects as irregular if they see asymmetric features, not taking into account if the object appears to be in a merger or not. This classification is therefore comparable to the asymmetry A.  Additionally, we use the merger classification. Classifiers are asked to identify any objects appearing to undergo interaction or showing a nearby companion. As such, this category is more prone to subjective interpretations by the classifiers and is explicitly not comparable to the asymmetry.

As expected from the similarities in the overall distributions, the rates of objects with high asymmetries (A $>$ 0.1) are not higher in the AGN hosts compared to control. When studying the rates as a function of X-ray luminosity, no differences between the AGN hosts and control are found. As a function of galaxy magnitude, there is a mild excess in high asymmetry rates at moderate galaxy masses; with AGN hosts having slightly higher asymmetries. However, this is not statistically significant. Similarly, we find that the rates of objects classified as irregular by human classifiers show no statistically significant difference between the AGN hosts and matched control. We also find that using a cut-off A$>$0.1 leads to similar overall asymmetry/irregularity rates as classification by humans.

We find the rates of human-classified mergers are significantly higher in the AGN hosts when compared to matched control. The excess rates for AGN compared to control seen in the merger rates are due to companions showing no signs of merger - not train-wreck mergers or even mergers showing some signs of interaction. A substantial fraction of these objects is associated with the cluster environment at z$\sim$0.75.

\section{Discussion}
\label{S:discussion}

We have compared the morphologies of X-ray selected AGN to those of matched galaxies of the same mass. The AGN studied are at redshift z=0.5-0.8 and have X-ray luminosities ($\log(L_{X} \ \textrm{[erg/s]}) \approx 41-44.5$). Assuming standard bolometric corrections \citep{runnoe_updating_2012,nemmen_quasar_2010}, the AGN luminosities span a range of about four orders of magnitude.

The host galaxies of moderately luminous X-ray selected AGN at low redshift show no strong differences in morphological properties (asymmetries, Sersic indices, ellipticities) compared to galaxies matched in mass or absolute H band magnitude. There is no statistically significant difference between AGN host galaxies and control galaxies in the rates of very high (A$>$0.1) asymmetry galaxies or the rates of galaxies classified as irregular by human classifiers. The asymmetry distributions and high asymmetry rates also do not show greater differences between AGN hosts and control as a function of either absolute galaxy magnitude or AGN X-ray luminosity. There are no differences between the asymmetries of X-ray obscured and unobscured AGN. The only statistically significant difference found between AGN and control galaxies is that AGN host galaxies located in a cluster environment show lower asymmetries as well as a higher skew, indicative of a stronger distribution tail at high asymmetries. Additionally, morphological inspections by human classifiers show that AGN hosts have higher rates of companions, although these show no signs of interaction. This difference is due to AGN found in the cluster environment.

Before comparing our results to those of other authors, we will briefly address the surface brightness limit and its influence on the types of merger features detected. We reach surface brightnesses around 26 mag/arcsec$^{2}$ in our study. Generally, the brightest merger features have surface brightnesses as high as 22 mag/arcsec$^{2}$, while most fainter tidal tails have surface brightnesses down to 25 mag/arcsec$^{2}$ \citep{elmegreen_smooth_2007}. Merger features are found even in local red and dead elliptical galaxies when depths of 28 mag/arcsec$^{2}$ are reached \citep{van_dokkum_recent_2005}. However, at these extremely faint surface brightness levels, merger features are also likely to trace disruption of small satellites \citep[e.g.][]{johnston_interpreting_2001}. Given the limits of our data, we are therefore likely to detect ongoing mergers, as well as tidal tails, while the data is not sensitive enough to detect disruption of smaller satellites. However, since we are primarily interested in the incidence of recent major mergers (mergers with galaxy mass ratios $\geq 0.1$ less than $\sim$ 1 Gyr in the past), this limit is ideal for the science goal of our study. The spatial resolution of our study is $\sim$0.5 kpc. We might therefore miss asymmetric features on very small scales.

We will start by comparing our results to previous studies of AGN host galaxy morphologies compared to control samples of quiescent galaxies. A number of studies have found no statistically significant results differences \citep{dunlop_quasars_2003,grogin_agn_2005,boehm_agn_2012,cisternas_bulk_2011}. \citet{dunlop_quasars_2003} studied samples of radio-loud and radio-quiet quasars at z$<$0.25 with luminosities comparable to those in our study. \citet{kocevski_candels:_2012} analysed human classifications of host galaxies of X-ray selected AGN at z$\approx$2 with slightly higher luminosities than those studied in this paper. \citet{cisternas_bulk_2011} examined the hosts of z$\sim$1 X-ray selected AGN. All used human classifiers to determine the incidence of merger features and found no differences when compared to control. \citet{grogin_agn_2005} and \citet{boehm_agn_2012} both used quantitative morphology measures on samples of X-ray selected AGN. \citet{grogin_agn_2005} studied a sample of AGN at redshift  (0.4 $<$ z $<$ 1.3) at the same rest wavelength as our study and \citet{boehm_agn_2012} at z$\sim$0.7 and slightly bluer wavelengths. Both studies found no differences between AGN hosts and quiescent galaxies. All these studies reach surface brightnesses where typical bright merger features should be visible \citep{elmegreen_smooth_2007}. They use similar depths of imaging observations as well as AGN luminosities as in this study and all but \citet{dunlop_quasars_2003} use X-ray selection. The rest-frame wavelength studied varies slightly between studies. This implies that our findings are typical of X-ray selected AGN. 

Very deep imaging studies have revealed differences between AGN and control galaxies in some cases. \citet{ramos_almeida_optical_2011} found that the host galaxies of radio-loud z$<$0.7 AGN have disturbances with significantly higher surface brightnesses than found in quiescent galaxies. \citet{ramos_almeida_optical_2011} interpret these signatures as signs of minor mergers or fly-by interactions in radio galaxies that are more recent in the AGN hosts than in the general galaxy population. \citet{bennert_evidence_2008} also found weak signs of interaction, consistent with either major or minor mergers about 1 Gyr in the past in a sample of local quasars. \citet{koss_merging_2010} studied low luminosity hard-Xray selected AGN using very deep imaging data and found stronger signs of interaction in AGN compared to quiescent galaxies. However, weak merger features can be due to major or minor mergers in the distant past or disruption of satellites. This makes interpretation of such findings challenging.

Other studies analysed the host galaxies of AGN without comparing to control samples. \citet{liu_active_2011} and \citet{letawe_study_2010} studied the host galaxies of low redshift (z$<$0.3) luminous quasars and found high incidences of merger signatures. \citet{canalizo_quasi-stellar_2001} studied a small sample of low redshift quasars with ULIRG-like FIR SEDs and luminosities and found high levels of disturbance in their host galaxies. However, those sources were selected to be IR luminous. Their host galaxy morphologies differ from the general quasar population that shows low levels of interaction \citep{bahcall_hubble_1997}.  \citet{urrutia_evidence_2008} studied a sample of high-redshift quasars with strong reddening and large incidence of outflows. Nearly 100\% of the red quasars show strong signs of recent interaction. These quasars are amongst the most luminous AGN in the Universe and are therefore not comparable to the AGN analysed in this study.

A lack of control samples makes interpretation of these findings difficult, in particular when quasars with ongoing starbursts are considered. \citet{veilleux_deep_2009} found that local PG quasars show relatively low incidence of strong merger features, while the sub-sample of PG quasars with ULIRG-like FIR luminosities show large incidence of merger features, consistent with the local ULIRG population. This can be explained if the main connection is between ULIRG-like FIR luminosities and mergers - rather than quasars and mergers.

Connections between AGN, starbursts and mergers have also been studied at higher redshift. \citet{juneau_widespread_2012} found that while there is no strong connection between AGN and high specific star formation rates, highly obscured AGN become more prevalent in star-bursting systems. \citet{kartaltepe_goods-herschel_2012} studied the incidence of mergers and AGN in high-redshift ULIRGs and found that both rise with increasing starburst luminosity. In both studies, AGN in these systems tend to be weak due to extreme levels of obscuration and not necessarily because they are intrinsically weak. These findings suggest that AGN are common in starburst/merger systems. However, this does not necessarily imply that all AGN are connected to starbursts and mergers. This agrees with the findings of \citet{ellison_galaxy_2013} that the AGN fraction is increased by a factor of almost four in post-merger systems when compared to control. This suggests that while mergers can trigger AGN \citep{ellison_galaxy_2013}, merger triggering does not dominate the AGN population (this study). In agreement with this picture, \citet{wild_timing_2010} find that AGN are associated with starbursts, but that this growth mode only plays a minor role in the local universe.

Finally, in disagreement with our study, \citet{treister_major_2012} combined data from the literature to study the incidence of major merger features and found that the merger fraction does rise with AGN luminosity. \citet{treister_major_2012} reached considerably higher AGN luminosities than our study, but did not use a control sample. Due to limited availability of literature studies, their sample is not uniformly selected and the high merger fraction at high AGN luminosities is caused by three studies with very different AGN properties and redshifts \citep{urrutia_evidence_2008,bahcall_hubble_1997,kartaltepe_goods-herschel_2012}. Part of the sample is selected through starburst features \citep{kartaltepe_goods-herschel_2012}. Due to limited information about the AGN in these starbursts, \citet{treister_major_2012} used the overall IR luminosity as a proxy for the AGN strength. It is therefore not clear if the results from \citet{treister_major_2012} will hold for uniformly selected samples over the same wide luminosity range.

\subsection{Do mergers matter?}

Before discussing the implications of our findings, we would first like to consider the statistical power of the study due to sample size.

Our sample is limited to 60 AGN. When binning in X-ray luminosity and absolute galaxy magnitude, we divide the sample into sub-samples of 10 objects each. We will therefore discuss the power for these two sample sizes. Due to the fact that much larger control samples are used, we are dominated by counting noise in the AGN sample.  The rate of high asymmetry objects in the full control sample (i.e. all control galaxies used in the study) is $\sim10\%$, consistent with the AGN sample. We therefore would like to know the \textit{excess} merger fraction above the intrinsic merger fraction of 10\% found in the control sample that our study is sensitive to.  For the full sample of all 60 AGN, a rate of 10/60 (i.e. six expected plus four excess) high asymmetry objects would be sufficient to reject the null hypothesis that the rate of objects with high asymmetry is identical to that in the control sample at a confidence p$<$0.05 (one tailed). This corresponds to only 6\% of the AGN population being connected to mergers not found in the general population. We can therefore state at 95\% confidence that less than 6\% of the AGN in the luminosity range studied are connected to mergers not found in the general population. For the analysis using bins in AGN luminosity and galaxy magnitude (10 object per bin), an intrinsic rate of 5/10 (i.e. one expected plus four excess) objects would be required to reject the null hypothesis with a confidence p$<$0.05 (one tailed). In each of the bins, we can therefore state that at 95\% confidence less than 40\% of AGN are connected to mergers not found in the general galaxy population. For the two sample comparisons between the asymmetry distributions, the power is less easily estimated since it will depend on the expected distributions of asymmetries for mergers, which is not known. The numbers for the high asymmetry rates also depend on the exact asymmetry value used as a cut-off for high asymmetry objects. However, these power estimations given here provide an estimate of the statistical power of this study.  To summarize, the sample size results in high enough power to reject the mergers for a large majority (94\%) as a likely trigger for the full sample as well as a majority (60\%) for the bins at a confidence of 95\%. 

A possible explanation for the lack of stronger merger signatures in AGN compared to control in our sample is a long delay between the merger event and AGN phase. Such a delay has been suggested by timing of starbursts and AGN activity. Theoretical models generally predict a delay of about 100 Myr between coalescence and the peak quasar activity \citep[see e.g.][]{hopkins_cosmological_2008,di_matteo_energy_2005}, which is a time-scale on which merger signatures should stay very apparent \citep{lotz_effect_2010,lotz_effect_2010-1}. \citet{wild_timing_2010} found that spectra of low luminosity AGN show signs of starbursts quenched several 100 Myr in the past. The presence of low surface brightness merger features in deep optical imaging data \citep{ramos_almeida_are_2011,ramos_almeida_optical_2011,bennert_evidence_2008}, in particular the presence of shells, which are indicative of a merger in the past, also speaks for a possible time-delay between merger and AGN activity. However, assuming typical delay times from several hundred Myr \citep{wild_timing_2010} up to $\sim$ 1 Gyr \citep{hyvnen_stellar_2007}, slightly elevated asymmetries in the host galaxies are still expected \citep{lotz_effect_2010,lotz_effect_2010-1}. We however do not observe such elevated levels of asymmetry.

Another possibility is that we might be missing the youngest, most obscured AGN that are closely connected to mergers. While dust obscuration in AGN is generally interpreted as a sign of the dusty torus, and therefore is indicative of orientation rather than an evolutionary phase, it is possible that there exists a sub-sample of AGN that are in a heavily buried phase where the obscuration is from host galaxy dust or gas, rather than the torus. These types of "buried" AGN could be connected to a very early phase of the merger-AGN life-cycle \citep{sanders_ultraluminous_1988}. Some observations indeed suggest that X-ray obscured AGN are prevalent in starburst system. Asymmetries in obscured and unobscured AGN in our sample are consistent, implying that this does not play a role for a large fraction of the X-ray selected AGN population. However, we note that the most heavily X-ray obscured AGN are missed by our selection method.

In this study, there appear to be only two differences between AGN host galaxies and matched control samples: a) in the cluster environment, host galaxies of AGN have lower asymmetries; b) AGN host galaxies have more nearby neighbours, particularly in the cluster environment. One possibility is a matching between AGN host galaxies to control galaxies located outside the cluster. While we choose a narrow redshift range for matching and reject matched galaxies far from the cluster area, there might still be a mismatch. It has been shown that galaxies in the centres of clusters show weaker signs of interaction \citep{adams_environmental_2012}. \citet{castellano_x-ray_2011} analysed the cluster studied here in detail and found it to be X-ray under-luminous. They suggested that AGN feedback in the past had removed most of the gas from the cluster. Such a scenario could also explain the lack of merger features in the AGN hosts: if AGN feedback removed the majority of cold gas from the cluster, galaxy interactions would be less likely to show tidal features during interactions.  The higher incidence of close-by companions is also consistent with merger triggering at an early stage of mergers, as found in many other studies \citep{koss_merging_2010,silverman_impact_2011,ellison_galaxy_2011,liu_active_2011,villforth_spectral_2012}. 

A final possibility is that the importance of mergers in AGN triggering is lower than expected, and other processes dominate over a wide range of AGN luminosities. Some theoretical models suggest that torques on galaxy wide scales \citep{angles-alcazar_black_2013,hopkins_analytic_2011} or violent disk instabilities driven by cold flows or other processes \citep[e.g.][]{dekel_cold_2009} could be drivers for AGN activity. Some observations suggests that clumpiness in disk galaxies correlates with AGN activity \citep[e.g.][]{bournaud_observed_2012}. However,clumpy disks are rare in our sample of galaxies.

\section{Summary \& Conclusions}
\label{S:summary}

We have presented a study aimed at understanding if and how the importance of merger triggering for AGN changes as a function of X-ray luminosity and host galaxy stellar mass. From theoretical considerations, it is expected that extreme processes such as major mergers are needed to trigger the most luminous AGN. In particular, assuming a typical quasar lifetime of $10^{8}$yr and AGN with $\log(L_{bol} \ \textrm{[erg/s]})=46$ requires as much as $2\times 10^{8} M_{\odot}$ of accreting material. This is comparable to about 1\% of typical gas masses in massive galaxies. Stripping such large gas masses likely requires extreme events. We present a quantitative analysis of the host galaxies of a sample of X-ray selected AGN with luminosities $41 < \log(L_{X} \ \textrm{[erg/s]}) < 44.5$ erg/s at low redshift ($0.5 < z < 0.8$). The host galaxies of AGN in our sample are compared to control galaxies matched in either stellar mass or absolute $F160W$ magnitude, which is found to trace stellar mass well. To ensure comparable morphology measures, we simulate 'fake AGN' by adding point sources with the AGN magnitude to control galaxies and consecutively removing them. We then compare the morphological properties of AGN hosts and control. Our findings can be summarized as follows.

\begin{itemize}
\item AGN hosts and control galaxies have asymmetry distributions consistent with the null hypothesis that the two samples are drawn from the same parent population. We also find no differences between AGN hosts and control galaxies asymmetries when binned in X-ray luminosity, even in the highest luminosity. The same is found when as a function of galaxy absolute magnitude.
\item There are no statistically significant differences in the fractions of high asymmetry objects, either  determined using asymmetry measures or from human classifier results. 
\item Both Sersic indices and ellipticities of AGN hosts and matched control galaxies are consistent with being drawn from the same parent population. Sersic indices are on average higher at higher galaxy masses but show only weak dependency on AGN luminosity.
\item We additionally test the hypotheses that merger triggering is most prevalent in an early obscured AGN phase. We divide the AGN sample into two sub-samples based on different levels of obscuration. Two-sample statistical test results are consistent with the null hypothesis that the two samples are drawn from the same parent population. We conclude that dust obscuration is unlikely to be a large effect in the incidence of mergers in our sample. Alternatively, if dust obscuration in an early merger phase plays a role, the percentage of AGN affected must be small.
\item The redshift range studied includes a cluster of galaxies at a redshift $z\sim0.75$ \citep{salimbeni_comprehensive_2009}. We find a difference between AGN hosts and control galaxies in the cluster at high statistical significance (p$<$0.01). AGN hosts in the cluster environment show on average \textit{lower} asymmetries, but \textit{higher} skew and kurtosis in the distribution of asymmetries when compared to control. However, it is unclear if this is due to a mismatch of AGN to galaxies not located in the cluster.  We also find higher incidences of nearby neighbours not showing interaction in the AGN hosts located in the cluster when compared to control. Both findings seem to indicate different processes in the cluster environment.
\item Given our sample size of 60 AGN and $\sim$700 control galaxies, we can infer with 95\% confidence that major mergers are responsible for $<$6\% of all AGN in our study as well as $<$40\% of the highest luminosity AGN in our sample ($\log(L_X \ \textrm{[erg/s]}) \sim 43.5$).
\item Our findings indicate that major mergers are either not the dominant triggering process at the AGN luminosities studied or the delay between the triggering and AGN activity is too long for significant merger features to remain apparent in the morphologies.
\end{itemize}

\section*{Acknowledgements}
We thank the anonymous referee for helpful comments. This work is based on observations taken by the CANDELS Multi-Cycle Treasury Program with the NASA/ESA HST, which is operated by the Association of Universities for Research in Astronomy, Inc., under NASA contract NAS5-26555. This work was supported in part by the USA National Science Foundation grant AST-1009628. Support for Program number HST-GO-12060 was provided by NASA through a grant from the Space Telescope Science Institute, which is operated by the Association of Universities for Research in Astronomy, Incorporated, under NASA contract NAS5-26555. Paola Santini would like to acknowledge funding through grant  ASI I/005/11/0. David Koo would like to acknowledge funding through grant NSF AST-0808133. Stephanie Juneau acknowledges financial support from the EC through an ERC grant StG-257720. For all statistical tests as well as estimation of probability density functions for statistical distribution, we use \textsc{SciPy} statistical packages\footnote{http://docs.scipy.org/doc/scipy/reference/stats.html}.

\bibliographystyle{mn2e}
\bibliography{hosts}

\appendix
\section{Measuring Unbiased Morphologies in AGN Hosts}
\label{S:appendix}

For this study, we measure the asymmetry of AGN hosts compared to their control galaxies. The central point sources in the AGN hosts are removed through PSF fitting, however, since this process might leave central residuals, we need to insure that this does not bias our morphological measures. The tests performed to ensure this will be summarized in this Appendix.

For this study, we use the asymmetry A, defined as:

\begin{equation}
A \equiv \sqrt{ \dfrac{ \sum \frac{1}{2} \times (I_{0} - I_{180})^{2} }{ I_{0} ^{2}}  }
\end{equation}

where $I_{0}$ is the image and $I_{180}$ is the image rotated by 180$^{\deg}$ \citep{conselice_relationship_2003}. For the purpose of this study, we use segmentation maps to avoid including noise from the background into the measurement. In the following, we will discuss different influences on the measured asymmetry. Centring is performed following \citep{conselice_asymmetry_2000}. We have ensured that the algorithm generally reaches a well-defined minimum, visual inspection is performed in addition. 

\subsection{AGN Contamination}

\begin{figure*}
\begin{center}
\includegraphics[width=5.5cm]{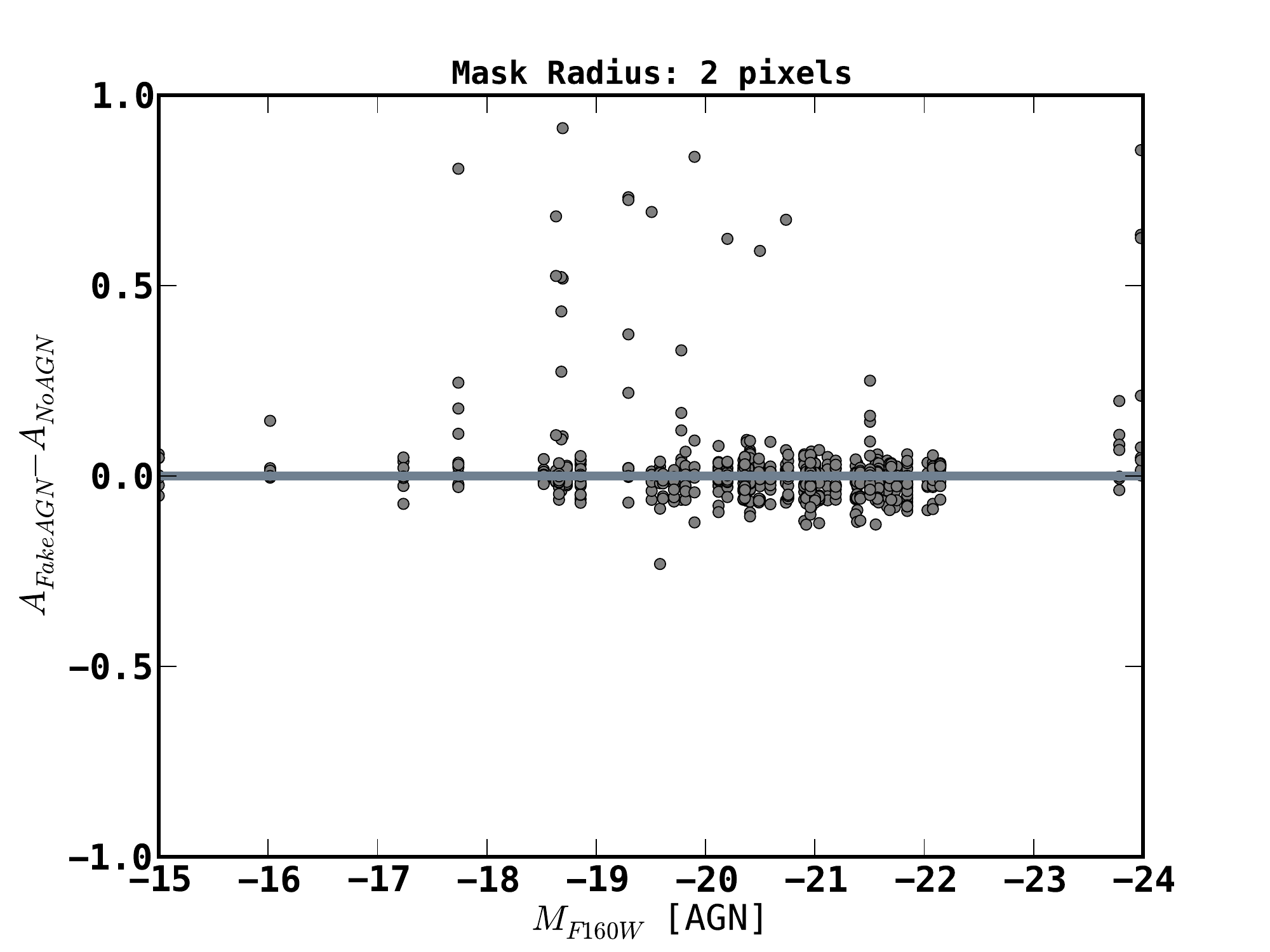}
\includegraphics[width=5.5cm]{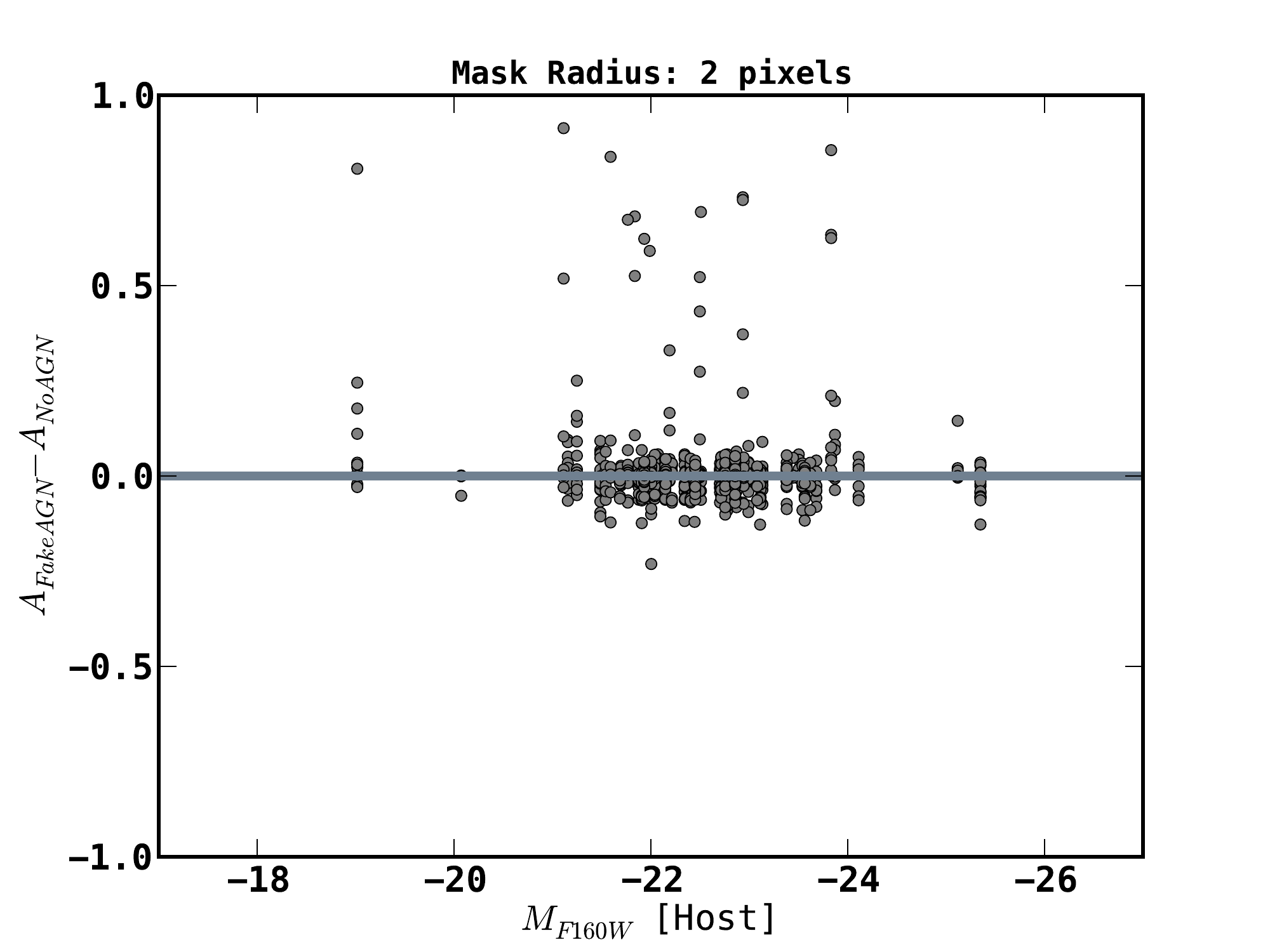}
\includegraphics[width=5.5cm]{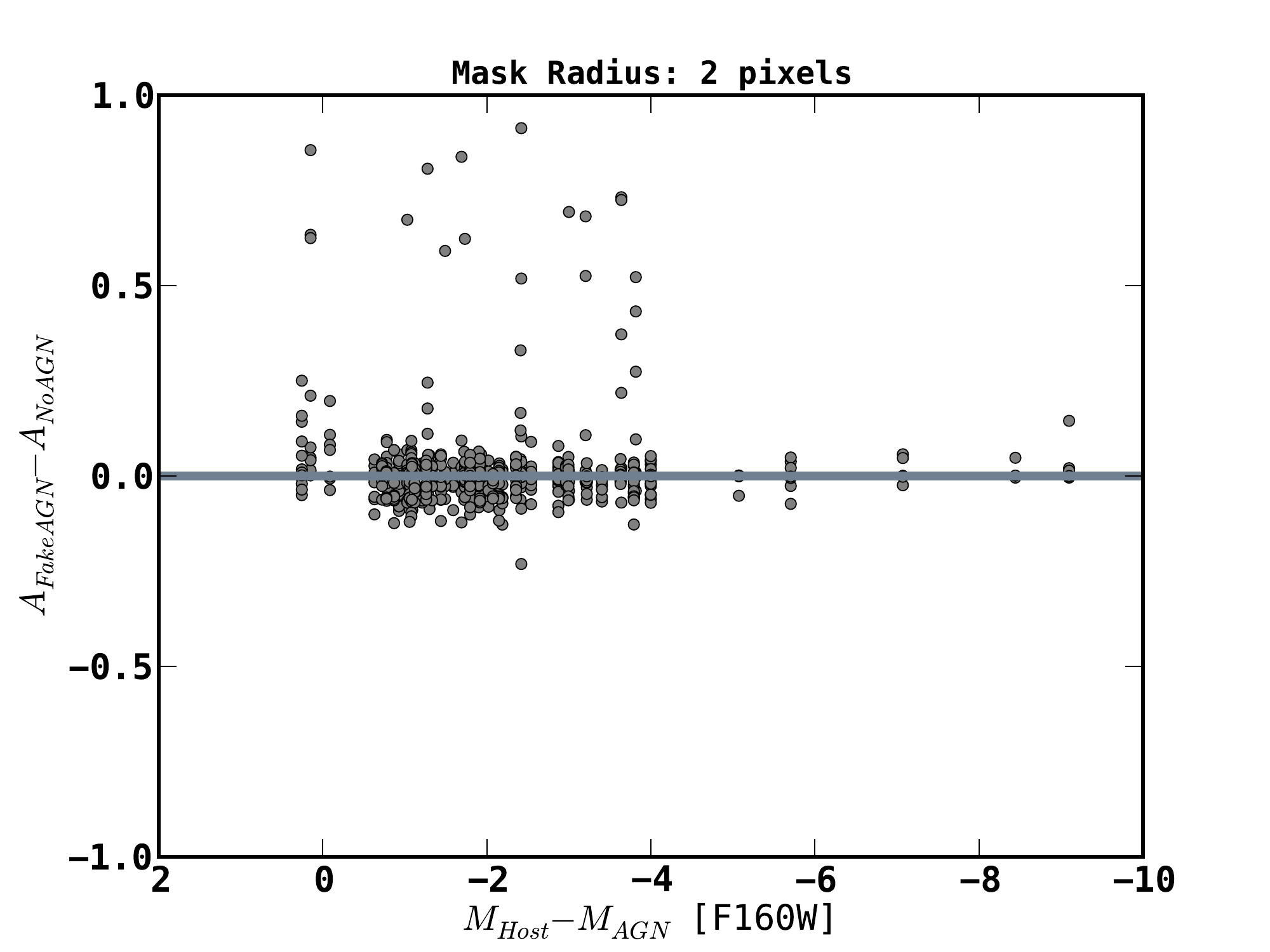}
\caption{Influence of AGN on asymmetry measures (2 pixel radius mask).}
\label{F:App_AGN}
\end{center}
\end{figure*}

While the adding of fake AGN in our study is aimed at simulating the influence of AGN on the asymmetry measures, we would still like to analyse the influence of the point source. For the control sample, we plot the difference between the asymmetry with and without adding of a Fake AGN in Fig. \ref{F:App_AGN}. As can be seen, the adding of moderate luminosity AGN does not greatly affect the asymmetry measures, and in the moderate range of AGN-galaxy contrast studied here, we do not find strong dependence on AGN magnitude.

\subsection{Masking radius used}

\begin{figure*}
\begin{center}
\includegraphics[width=5.5cm]{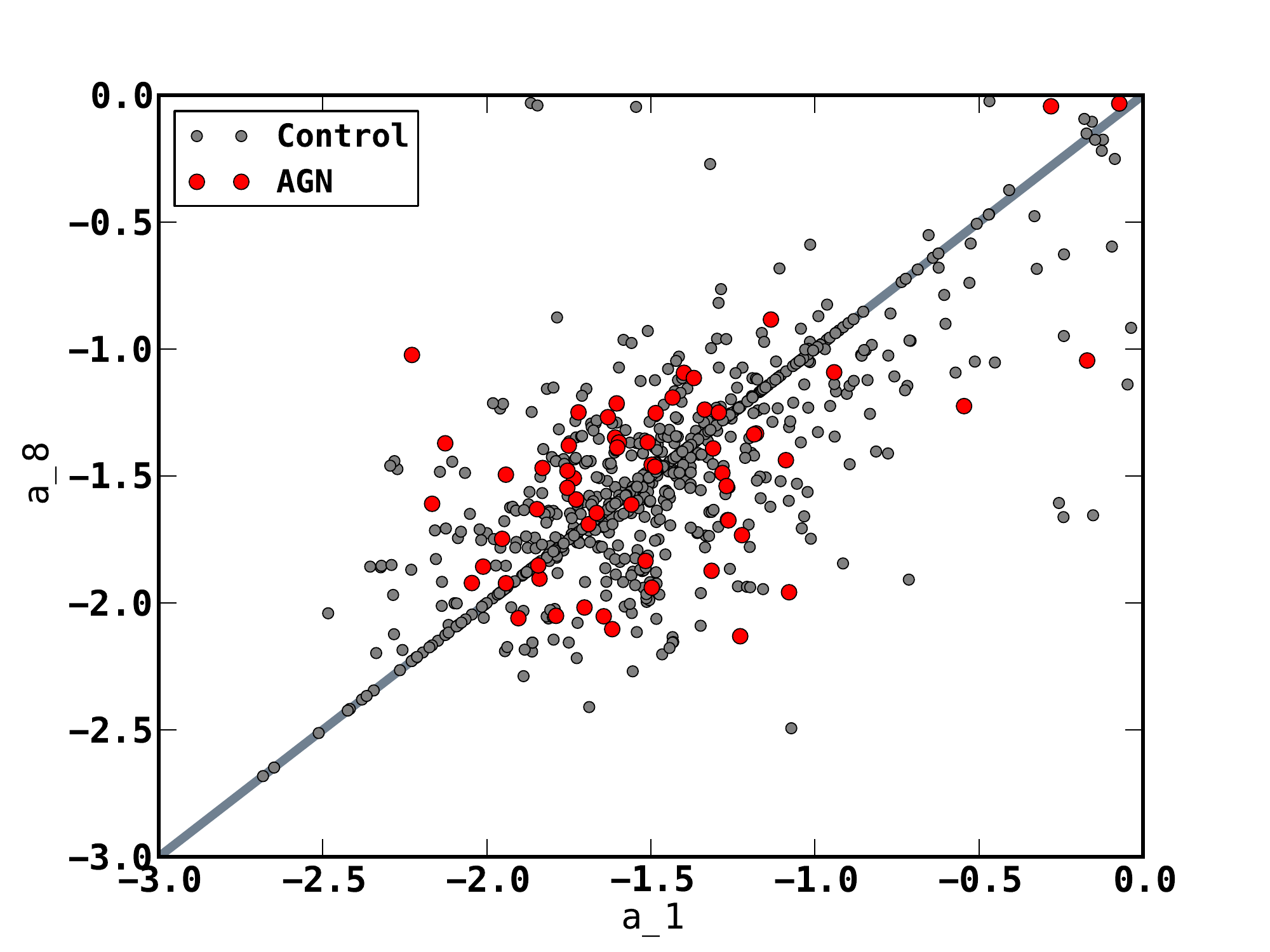}
\includegraphics[width=5.5cm]{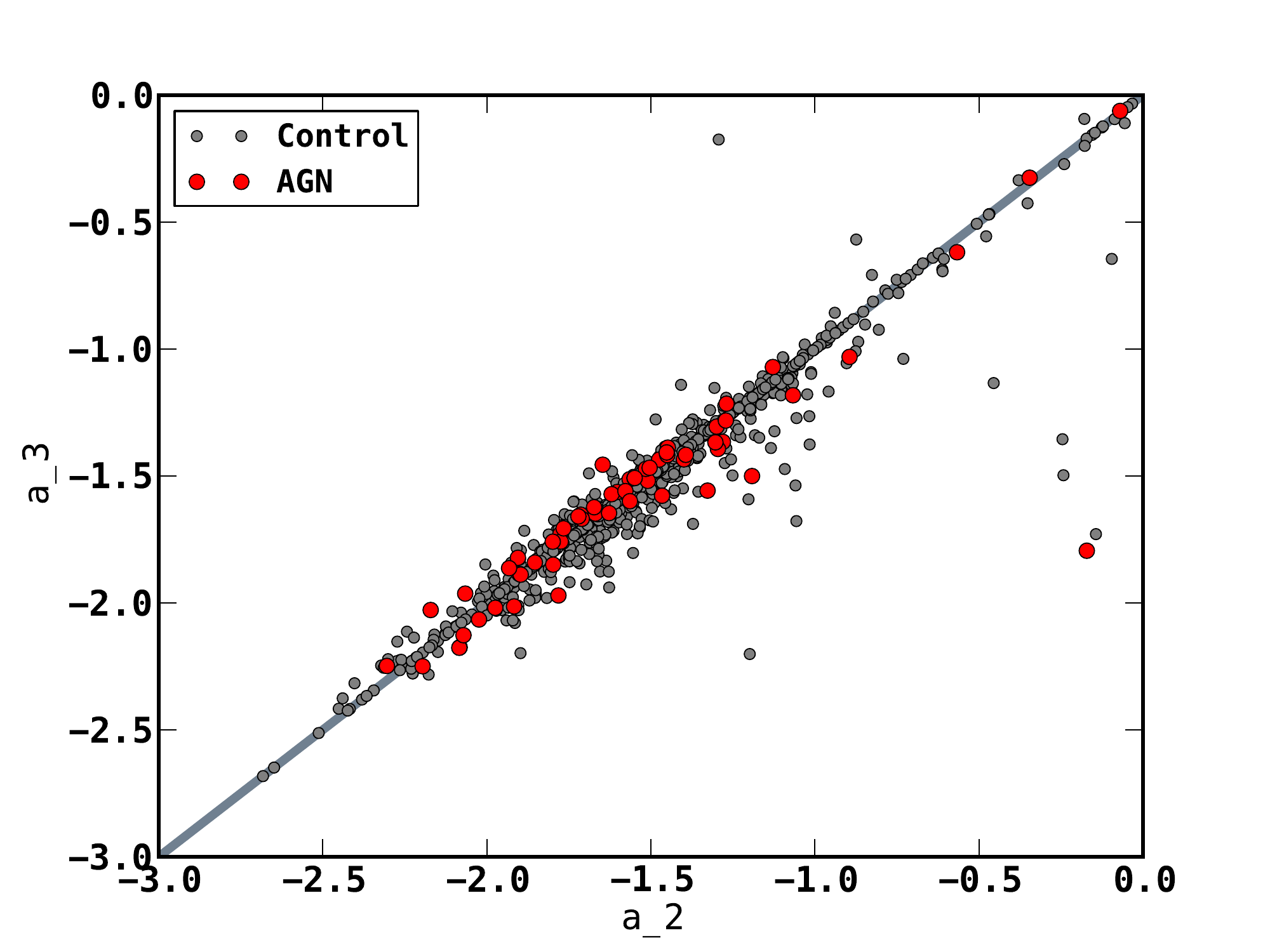}
\includegraphics[width=5.5cm]{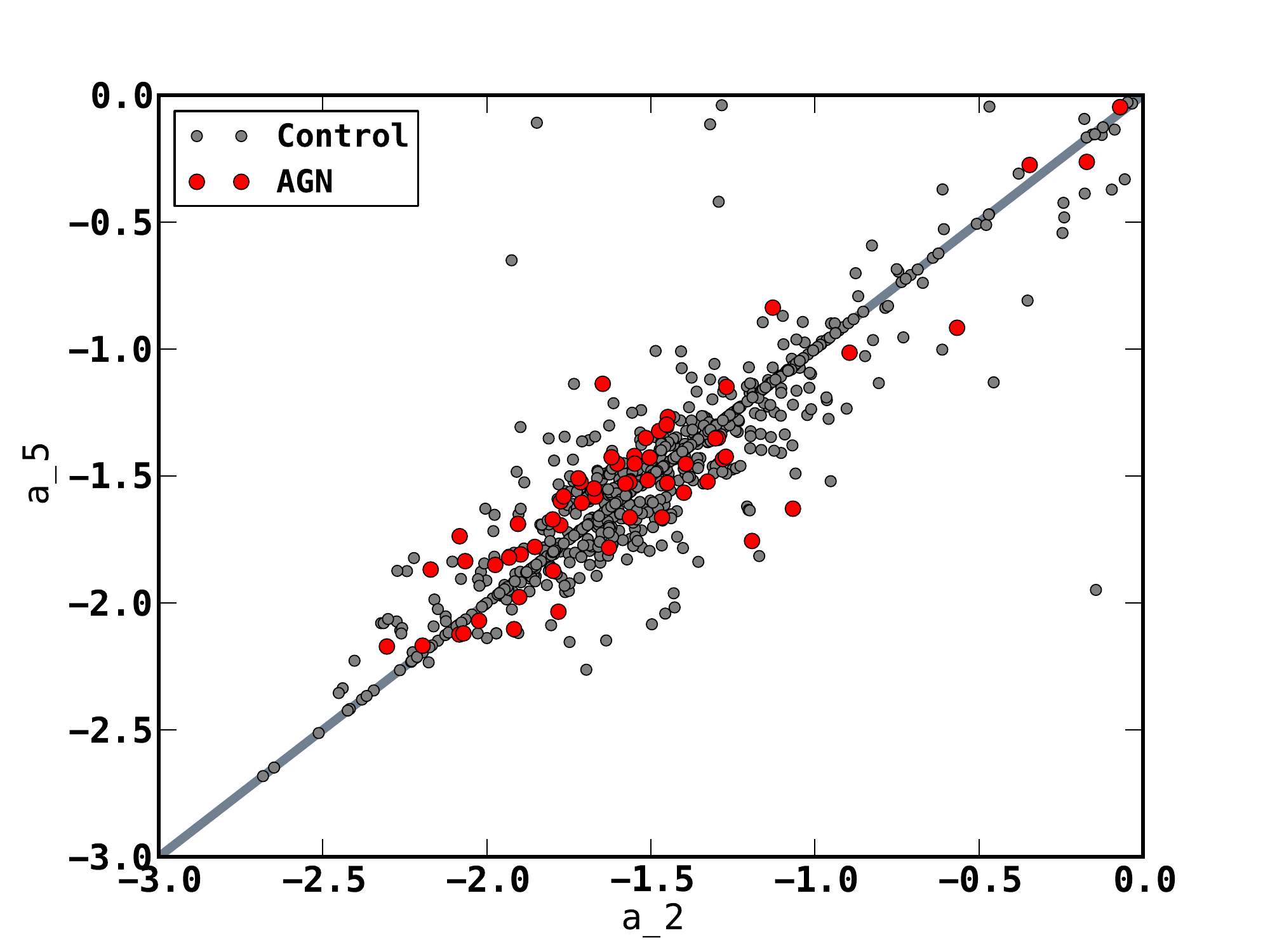}
\caption{Influence of mask size on asymmetry.}
\label{F:App_mask}
\end{center}
\end{figure*}

Secondly, we would like to shortly discuss the influence of mask size on the asymmetry values. In Figure \ref{F:App_mask}, we show comparisons between different mask sizes with radii between 1 pixel and 8 pixels. As expected, the values correlate well when the radii are similar, and greater differences are seen when the radii are very different, this is well expected. After visual inspection, we choose a radius of 2 pixel since it covers most corrupted pixels in the higher luminosity AGN.

Checks of the results concerning differences between the AGN hosts and control galaxies however show that the masking radius does not alter the overall results of the study.

\section{Full Statistical Properties of Sample}
\label{S:appendix_stats}

\begin{table*}
\begin{minipage}{180mm}
\caption{Statistical properties of host galaxy morphology measures for AGN hosts as well as matched control. The values for the control sample are derived by bootstrapping and the mean value and standard deviation from the bootstrapping are given. First column: morphological property, second column: AGN property used for binning; third column: mean of AGN property used for binning; $\mu$: mean value of morphological measure (AGN and control); $\sigma$: standard deviation of morphological measure for AGN and control; skew: skew of morphological measure for AGN and control; kurtosis: kurtosis of morphological measure for AGN and control.}
\begin{tabular}{@{}ccccccccccc@{}}
\hline
& & & \multicolumn{2}{c}{$\mu$} & \multicolumn{2}{c}{$\sigma$} & \multicolumn{2}{c}{skew} &  \multicolumn{2}{c}{kurtosis} \\
 &  & $\mu$ & AGN & Control & AGN & Control & AGN & Control & AGN & Control\\
\hline
log(A) & $log(L_{Xray})$ (All) & 42.27 & 0.14 & 0.06 $\pm$ 0.00 & 0.15 & 0.12 $\pm$ 0.01 & 4.15 & 5.14 $\pm$ 0.51 & 19.97 & 31.87 $\pm$ 6.98\\ 
log(A) & $log(L_{Xray})$ & 41.18 & 0.14 & 0.07 $\pm$ 0.01 & 0.25 & 0.13 $\pm$ 0.03 & 2.31 & 3.76 $\pm$ 0.74 & 6.81 & 17.32 $\pm$ 6.84\\ 
log(A) & $log(L_{Xray})$ & 41.59 & 0.03 & 0.07 $\pm$ 0.02 & 0.03 & 0.15 $\pm$ 0.04 & 2.08 & 4.01 $\pm$ 0.88 & 6.14 & 19.60 $\pm$ 8.14\\ 
log(A) & $log(L_{Xray})$ & 41.91 & 0.09 & 0.06 $\pm$ 0.01 & 0.20 & 0.12 $\pm$ 0.04 & 2.63 & 4.60 $\pm$ 1.18 & 7.98 & 26.56 $\pm$ 10.40\\ 
log(A) & $log(L_{Xray})$ & 42.33 & 0.03 & 0.04 $\pm$ 0.00 & 0.01 & 0.03 $\pm$ 0.00 & 0.13 & 1.49 $\pm$ 0.22 & 2.01 & 4.66 $\pm$ 0.96\\ 
log(A) & $log(L_{Xray})$ & 43.04 & 0.06 & 0.04 $\pm$ 0.01 & 0.13 & 0.06 $\pm$ 0.03 & 2.61 & 3.41 $\pm$ 1.97 & 7.93 & 19.33 $\pm$ 15.38\\ 
log(A) & $log(L_{Xray})$ & 43.60 & 0.04 & 0.08 $\pm$ 0.01 & 0.02 & 0.14 $\pm$ 0.03 & 0.78 & 3.94 $\pm$ 0.76 & 2.50 & 19.47 $\pm$ 7.15\\ 
& & & & & & & & & &\\ 
log(A) & $M_{Host}$ (All) & -22.53 & 0.06 & 0.06 $\pm$ 0.00 & 0.15 & 0.12 $\pm$ 0.01 & 4.15 & 5.14 $\pm$ 0.49 & 19.97 & 31.81 $\pm$ 6.61\\ 
log(A) & $M_{Host}$ & -24.02 & 0.04 & 0.07 $\pm$ 0.01 & 0.02 & 0.12 $\pm$ 0.04 & 0.81 & 4.16 $\pm$ 1.06 & 2.35 & 22.60 $\pm$ 9.23\\ 
log(A) & $M_{Host}$ & -23.10 & 0.09 & 0.05 $\pm$ 0.01 & 0.14 & 0.09 $\pm$ 0.04 & 1.78 & 4.75 $\pm$ 1.56 & 4.53 & 28.62 $\pm$ 12.48\\ 
log(A) & $M_{Host}$ & -22.75 & 0.09 & 0.04 $\pm$ 0.01 & 0.20 & 0.07 $\pm$ 0.03 & 2.63 & 4.38 $\pm$ 1.99 & 8.00 & 26.97 $\pm$ 14.44\\ 
log(A) & $M_{Host}$ & -22.32 & 0.10 & 0.05 $\pm$ 0.01 & 0.25 & 0.07 $\pm$ 0.02 & 2.66 & 3.86 $\pm$ 1.05 & 8.09 & 20.36 $\pm$ 8.53\\ 
log(A) & $M_{Host}$ & -21.94 & 0.02 & 0.07 $\pm$ 0.01 & 0.01 & 0.13 $\pm$ 0.03 & 0.00 & 3.95 $\pm$ 0.91 & 1.56 & 18.87 $\pm$ 8.53\\ 
log(A) & $M_{Host}$ & -21.06 & 0.05 & 0.08 $\pm$ 0.02 & 0.03 & 0.16 $\pm$ 0.04 & 1.21 & 4.05 $\pm$ 0.85 & 3.94 & 20.38 $\pm$ 8.00\\ 
\hline
\end{tabular}
\label{T:stats}
\end{minipage}
\end{table*}

\label{lastpage}

\end{document}